%% file: ms.tex
\pdfoutput=1 

\documentclass[a4paper, 11pt, final]{article}

\input{ms_a_preamble}

\begin{document}


\renewcommand{\figureautorefname}{Fig.}
\onehalfspacing


\input{ms_b_frontmatter}



\newpage

\input{1-Intro}
\input{2-Background}

\input{3-SurvivalModelling}

\input{4-Conclusion}


\appendix
\input{5-Branch_PD-Reuse}

\input{6-Diagnostics}

\input{7-InputSpace}


\singlespacing
\printbibliography 
\onehalfspacing

\section{Statements and declarations}
\subsection*{Funding declaration}
\noindent This work is financially supported wholly/in part by the National Research Foundation of South Africa (Grant Number 126885).
%
\subsection*{Competing interest declaration}
The authors have no relevant financial or non-financial interests to disclose

\subsection*{Author contributions}
All authors contributed to the study conception and design. Material preparation, data collection, and analysis were performed by Dr Arno Botha. All authors have read and approved the final manuscript.



\end{document}

%% file: ms_a_preamble.tex
\usepackage{amssymb} 
\usepackage{amsthm} 
\usepackage{amsmath,empheq}
\usepackage{bbm}

\DeclareMathAlphabet{\mathcal}{OMS}{cmsy}{m}{n}
\DeclareMathAlphabet\mathbfcal{OMS}{cmsy}{b}{n}

\DeclareFontFamily{U}{dutchcal}{\skewchar\font=45 }
\DeclareFontShape{U}{dutchcal}{m}{n}{<-> s*[1.0] dutchcal-r}{}
\DeclareFontShape{U}{dutchcal}{b}{n}{<-> s*[1.0] dutchcal-b}{}
\DeclareMathAlphabet{\mathcald}{U}{dutchcal}{m}{n}
\SetMathAlphabet{\mathcald}{bold}{U}{dutchcal}{b}{n}
\DeclareMathAlphabet\mathcalz{T1}{pzc}{mb}{it}

\usepackage[nice]{nicefrac} 
\usepackage{siunitx} 

\usepackage{newtxtext, newtxmath} 

\usepackage{anyfontsize}
\usepackage[utf8]{inputenc}
\usepackage[T1]{fontenc}
\usepackage{microtype} 

\providecommand{\JEL}[1]{\textit{\textbf{JEL: }} #1}
\providecommand{\keywords}[1]{\textbf{\textit{Keywords--- }} #1}

\usepackage{setspace} 

\usepackage{parskip} 
\usepackage{etoolbox}
\usepackage{titlesec}
\titleformat{\section}{\normalfont\Large\bfseries}{\thesection}{1em}{}
\titleformat{\subsection}{\normalfont\large\bfseries}{\thesubsection.}{1em}{}
\titleformat{\subsubsection}{\normalfont\normalsize\itshape}{\thesubsubsection.}{1em}{}

\usepackage{abstract}

\renewenvironment{abstract}
 {\normalfont
  \begin{center}
  \bfseries \abstractname\vspace{-.5em}\vspace{0pt}
  \end{center}
  \list{}{
    \setlength{\leftmargin}{0cm}%
    \setlength{\rightmargin}{\leftmargin}%
  }%
  \item\relax}
 {\endlist}

\usepackage{authblk} 
\usepackage[bottom]{footmisc} 

\usepackage{multicol} 

\usepackage{enumitem} 

\usepackage{graphicx} 
\usepackage{float} 
\usepackage{subcaption}
\usepackage{afterpage} 

\usepackage{printlen}

\usepackage[labelfont=bf]{caption}
\usepackage{color, colortbl}
\definecolor{LightGray}{rgb}{0.93,0.914,0.914}    
\usepackage{soul} 
\usepackage{longtable,rotating} 
\usepackage{booktabs} 
\usepackage{multirow} 
\usepackage{arydshln} 
\usepackage{bigdelim} 

\usepackage[most]{tcolorbox}            

\PassOptionsToPackage{hyphens}{url} 

\usepackage{fancyhdr}
\fancyheadoffset{0cm}
\makeatletter
\newcommand{\quickwordcount}[1]{
  \immediate\write18{texcount -quiet -incbib -sub=none -utf8 -1 -sum -merge -encoding=utf8 #1.tex > #1-words}%
  \immediate\openin\somefile=#1-words
  \read\somefile to \@@localdummy
  \immediate\closein\somefile
  \setcounter{wordcounter}{\@@localdummy}
  \@@localdummy
}
\makeatother

\usepackage{silence}
\usepackage[style=apa,backend=biber,natbib,hyperref]{biblatex}
\DeclareLanguageMapping{british}{british-apa}
\defbibenvironment{bibliography}{\enumerate}{\endenumerate}{\item}

\usepackage[colorlinks=true,allcolors=gray]{hyperref} 
\let\orgautoref\autoref

\renewcommand{\autoref}[1]
{%
\def\equationautorefname{Eq.}%
\def\sectionautorefname{Sec.}%
\def\subsectionautorefname{Subsec.}%
\def\figureautorefname{Fig.}%
\def\subfigureautorefname{Fig.}%
\orgautoref{#1}%
}

\usepackage[nameinlink,capitalise]{cleveref}

\usepackage{algorithm,algpseudocode}

\makeatletter
\newlength{\trianglerightwidth}
\algnewcommand{\LineCommentCont}[1]{\Statex \hskip\ALG@thistlm%
  \parbox[t]{\dimexpr\linewidth-\ALG@thistlm}
{\leftskip=\algorithmicindent
  \hangindent=\algorithmicindent 
  \hangafter=1%
  \strut\makebox[\algorithmicindent][c]{$\triangleright$}#1\strut}
  } 
\makeatother

\usepackage[left=1.8cm, right=1.8cm, bottom=2.5cm, top=2.5cm]{geometry}

\usepackage{listings}
\usepackage{xcolor}

%% file: ms_b_frontmatter.tex

\newcommand{\MainTitleText}{Approaches for modelling the term-structure of default risk under IFRS 9: A tutorial using discrete-time survival analysis}

\title{\fontsize{20pt}{0pt}\selectfont\textbf{\MainTitleText
}}


\author[,a,b]{\large Arno Botha \thanks{ ORC iD: 0000-0002-1708-0153; email: \url{arno.spasie.botha@gmail.com}}}
\author[,a,b]{\large Tanja Verster \thanks{ ORC iD: 0000-0002-4711-6145; Corresponding author: \url{tanja.verster@nwu.ac.za}}}
\affil[a]{\footnotesize \textit{Centre for Business Mathematics and Informatics \& Unit for Data Science and Computing, North-West University, Potchefstroom, South Africa}}
\affil[b]{\footnotesize \textit{National Institute for Theoretical and Computational Sciences (NITheCS), Potchefstroom, South Africa}}
\renewcommand\Authands{, and }

    

\makeatletter
\renewcommand{\@maketitle}{
    \newpage
     \null
     \vskip 1em%
     \begin{center}%
      {\LARGE \@title \par
      	\@author 
        \par}
     \end{center}%
     \par
 } 
 \makeatother
 
 \maketitle

{
    \setlength{\parindent}{0cm}
    \rule{1\columnwidth}{0.4pt}
    \begin{abstract}
        Under the International Financial Reporting Standards (IFRS) 9, credit losses ought to be recognised timeously and accurately. This requirement belies a certain degree of dynamicity when estimating the constituent parts of a credit loss event, most notably the probability of default (PD). It is notoriously difficult to produce such PD-estimates at every point of loan life that are adequately dynamic and accurate, especially when considering the ever-changing macroeconomic background. In rendering these lifetime PD-estimates, the choice of modelling technique plays an important role, which is why we first review a few classes of techniques, including the merits and limitations of each. Our main contribution however is the development of an in-depth and data-driven tutorial using a particular class of techniques called discrete-time survival analysis. This tutorial is accompanied by a diverse set of reusable diagnostic measures for evaluating various aspects of a survival model and the underlying data. A comprehensive R-based codebase is further contributed. We believe that our work can help cultivate common modelling practices under IFRS 9, and should be valuable to practitioners, model validators, and regulators alike.
    \end{abstract}
     
    \keywords{IFRS 9; Credit risk; Term-structure; Survival analysis; Discrete-time; Time-dependent diagnostics.}
     
     \JEL{C33, C41, C52, G21.}
    
    \rule{1\columnwidth}{0.4pt}
}

\noindent Word count (excluding front matter and appendices): 12809 


%% file: 1-Intro.tex
\section{Introduction}
\label{sec:intro}

The International Financial Accounting Standard (IFRS) 9 from the \citet{ifrs9_2014} brought about a paradigm shift in the modelling of credit risk, i.e., the loss associated with borrowers who may default on their loans. IFRS 9 posits that the value of a financial asset ought to be adjusted over time in lieu of the asset's underlying credit risk. 
The principle is to forfeit a fraction of a bank's income today, and to do so regularly, in offsetting the written-off loan of tomorrow. Doing so would smooth away most of the volatility in earnings over time, thereby modulating abrupt spikes in the value of write-offs. In turn, the base uncertainty of losses is rendered into a more predictable construct, which is a central tenet of risk management, as discussed by \citet[pp.~38--4]{VanGestel2009book}.
The forfeited amount is funnelled into a loss provision, which should be regularly updated using a statistical model of the asset's \textit{expected credit loss} (ECL). This ECL-model represents the probability-weighted sum of cash shortfalls that are expected to be lost over a certain time horizon. From \citet[\S 5.5.17--18, \S B5.5.28--31, \S B5.5.41--44]{ifrs9_2014}, this ECL-amount should be estimated without bias and across a range of possible outcomes that can affect the asset's value over its lifetime. This estimation task ought to consider the time value of money, past events, current conditions, and forecasts of future economic conditions.
In so doing, the asset's value is comprehensively updated at each reporting date (usually monthly), either by raising more from earnings, or by releasing a portion of the provision back into the income statement.

\begin{figure}[ht!]
\centering\includegraphics[width=1\linewidth,height=0.3\textheight]{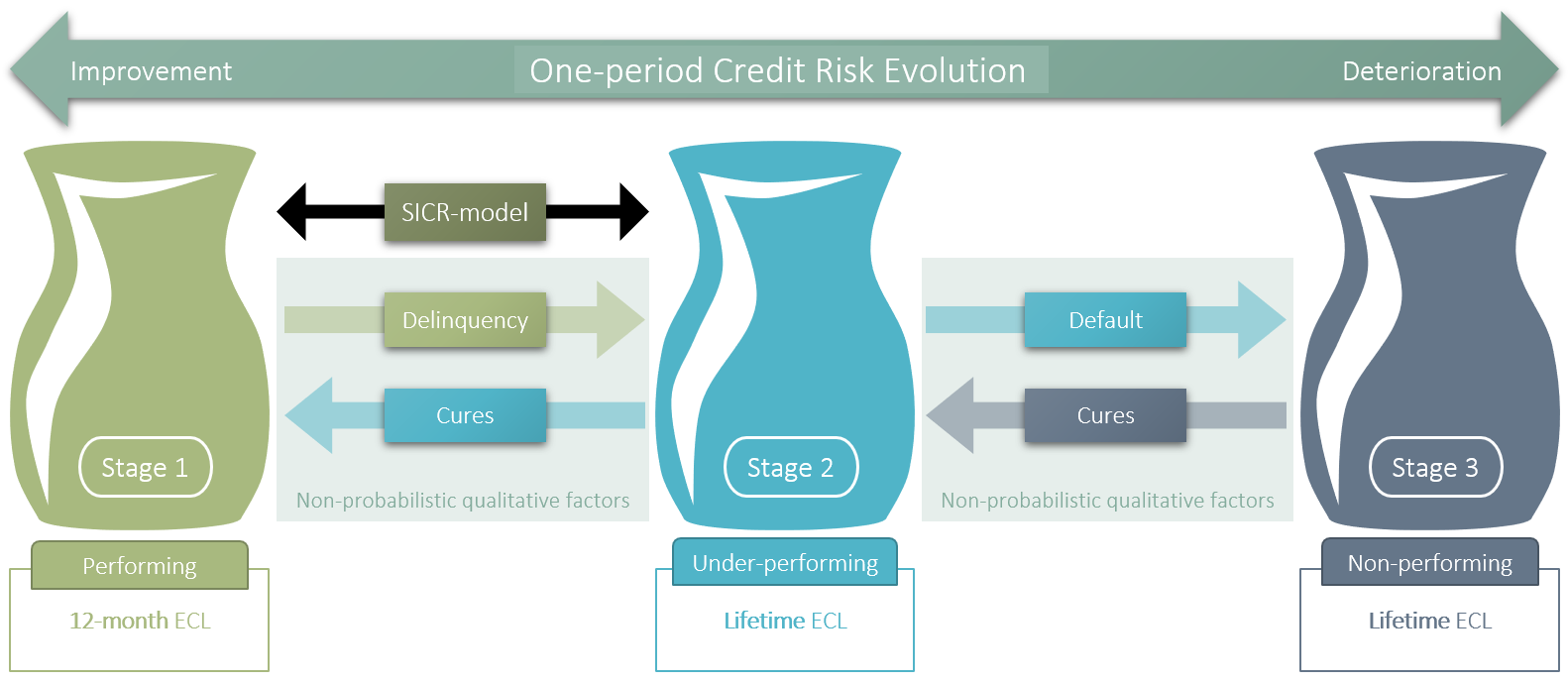}
\caption{Illustrating the one-period evolution of credit risk and its deterioration under IFRS 9, which constitutes the staged impairment of a loan. Each subsequent stage implies greater deterioration and therefore a larger ECL-estimate. Arrows indicate possible migrations, subject to meeting certain qualitative criteria. The exception is the probabilistic SICR-component (shaded in dark green), which can include various factors in predicting a SICR-event. From \citet{botha2025sicr}.}\label{fig:IFRS9_Stages}
\end{figure}

IFRS 9 further requires a staged approach for estimating the ECL-amount, as stated in \S 5.5.3--5.5.5 in IFRS 9, where this approach is based on the degree of deterioration in the asset's credit risk. As discussed by \citet{botha2025sicr} and illustrated in \autoref{fig:IFRS9_Stages}, each stage requires a greater ECL-amount in reflecting a general pattern of decay (or improvement) in credit quality over time. Most loan assets would typically reside in Stage 1, which have not experienced a \textit{significant increase in credit risk}, or a SICR-event. Conversely, Stage 2 includes those underperforming loans that have experienced such a SICR-event, though cannot yet be classified as fully credit-impaired or in default; Stage 2 is therefore a middle ground of sorts. Those loans with objective evidence of impairment are classified into Stage 3, which belies the fact that their future cash flows are likely compromised. IFRS 9 further differentiates these stages by requiring the ECL-value to be estimated over 12 months for Stage 1, and over the asset's remaining lifetime for Stages 2-3. Put differently, a first-stage loss is the portion of the lifetime ECL-amount that is lost over the next 12 months, whereas a second-stage (or third-stage) loss should consider all possible loss-inducing events over the entirety of remaining loan life. For greater detail, kindly consult the works of \citet{EY2014}, \citet{PWC2014}, and \citet[\S 1]{bellini2019}.
In this work, we shall limit our review to methods for estimating parts of the ECL over the entire lifetimes of loans, largely since these methods should already cater for ELC-estimation over the next 12 months (Stage 1).

The main thrust of IFRS 9 is to recognise credit losses more timeously and more accurately, which suggests a certain degree of dynamicity in the ECL-estimation process. In modelling the ECL, \citet{skoglund2017} explained that this process depends on a few risk parameters, of which the most important one is perhaps the \textit{probability of default} (PD), or default risk. Estimating a borrower's PD involves finding a statistical relationship between a set of input variables and the binary-valued repayment outcome (i.e., defaulted or not) over some outcome period.
In rendering dynamic PD-estimates under IFRS 9, risk models should have the ability to project default risk over a variety of time horizons across loan life, particularly given the ever-changing macroeconomic background. By implication, the underlying risk model should produce marginal PD-estimates as a function of a rich set of input variables, preferably including macroeconomic covariates. These marginal PD-estimates are produced at each discrete period $t=t_1,\dots,\mathcal{T}$ during a loan's lifetime $\mathcal{T}$, starting from its time of initial recognition $t_1$. We shall call the collection of these time-dependent PD-estimates the \textit{term-structure} of default risk.

However, rendering such dynamic and time-dependent PD-estimates is fraught with challenges. One aspect hereof is the fact that `default' is not necessarily an absorbing state into which a loan is forever trapped, as discussed by \citet[pp.~73-83]{botha2021phd} and empirically explored by both \citet{botha2025multistate} and \citet{botha2025recurrentEvents}. A loan may cure from the default state, become subject to default risk again, and default once again; constituting a repeatable cycle of recurrent events. One may therefore view `default' as a transient state, thereby leveraging the full credit histories that may otherwise be replete with such recurrent defaults.
This dynamicity is recognised in certain regulations, which require banks to grade loans as performing whenever default criteria ceases to apply. We refer the interested reader to \S 36.74 of the Basel framework from the \citet{basel2019}, and to Article 178(5) of the Capital Requirements Regulation (CRR), as promulgated by the \citet{eu2013CRR} for the EU-market.
Another major challenge is mired in the fact that `default' is not the only failure-inducing event. A loan may experience prepayment (or settlement), write-off, or restructure; all of which can affect the risk of loss under IFRS 9. These \textit{competing risks} will preclude the loan from defaulting, as well as impact the size of the risk set over time.
Overall, the underlying credit risk models should duly cater for this dynamicity over loan life in producing time-dependent PD-estimates that are suitably accurate.

In this work, and as part of our contributions, we shall review and synthesise the latest advances in PD term-structure modelling, which can address some of the aforementioned challenges. 
That said, other authors have already reviewed some of these approaches in other works of literature. In fact, and as we shall demonstrate later, most of these works cover a large breadth of broad modelling approaches and techniques. Our review shall instead focus on depth and explore a chosen class of techniques, with a particular focus on formulation, implementation, model diagnostics, and practicability.
We also found that these other reviews are somewhat lacking regarding the aforementioned aspects, at least when considering all of these aspects together. Some reviews mixed the various modelling methods across wholesale and retail lending, which can differ greatly in data availability and even the structure thereof; all of which has practical implications for the choice of method.
As such, and as our main contribution, this work shall aim towards crafting a complete and in-depth tutorial on modelling the term-structure of default risk, using a particular class of techniques known as \textit{survival analysis} with retail credit data. We also contribute a rich R-based codebase, which is available publicly.
This tutorial should be of considerable value to practitioners, model validators, and regulators alike in helping to shape common modelling practice under IFRS 9. Moreover, our tutorial bears some novelty in being the first such work in literature, at least to the best of the authors' knowledge.

This paper is structured as follows. In \autoref{sec:background}, we shall broadly review the available modelling approaches, thereby forming a taxonomy in which our own work can be positioned.
As our main focus, we shall explore in \autoref{sec:survival} the use of discrete-time survival analysis in deriving lifetime PD-estimates. Aspects hereof include: 1) formulating such models; 2) showcasing the underlying data structures using standardised notation within the credit risk context; 3) providing guidance regarding model fitting and implementation; and 4) presenting four diagnostic measures for evaluating the resulting survival models. These aspects therefore culminate in a comprehensive and data-driven modelling tutorial. In particular, this work leverages residential mortgage data from a large and representative South African bank in demonstrating the various modelling steps and powering the associated diagnostics. It is accompanied by a rich R-based codebase of \citet{botha2025termStructureSourcecode}, which has reusable components and functions; see the appendix. This codebase can be easily adapted for use within other real-world IFRS 9 modelling spaces.
The tutorial is finally concluded in \autoref{sec:conclusion} and some general recommendations are provided. 
Regarding ancillary material, we scrutinise in \autoref{app:PD_reuse} another popular class of techniques for deriving lifetime PD-estimates from existing PD-models. This class of techniques is reformulated using new and standardised notation, and we have reviewed these techniques critically in highlighting the dangers of their pursuit, despite their inherent simplicity.
Other than the tutorial, we also contribute in \autoref{app:tDiagnostics} a succinct review of time-dependent model diagnostics, which are newly formulated within the context of credit risk modelling, and used in \autoref{sec:survival_diagnostics}.

%% file: 2-Background.tex
\section{A broad review of approaches to deriving lifetime PD-estimates}
\label{sec:background}

In addressing the aforementioned challenges (recurrent defaults \& competing risks), some authors have already reviewed a few appropriate modelling techniques for producing dynamic lifetime PD-estimates under IFRS 9. These reviews are critically discussed in \autoref{sec:background_reviews}, including the merits and limitations of each. Most of the reviews ultimately recommend the use of survival analysis in modelling the lifetime PD, which is why we review the literature on this class of techniques in \autoref{sec:background_surv} within the ambit of credit risk modelling. Finally, we survey in \autoref{sec:background_other} a few other methods in producing lifetime PD-estimates in the interest of completeness.

\subsection{Surveying previous reviews of methods for lifetime PD-estimation}
\label{sec:background_reviews}

As a start, \citet{skoglund2017} briefly outlined an approach for scaling existing PD-models under Basel towards an IFRS 9 environment. This approach is part of a broader class of approaches that re-uses existing models, which we shall critically review in \autoref{app:PD_reuse}. Thereafter, the author discussed a Cox-regression modelling strategy, and considered a multistate Markov-type modelling setup to generate the PD term-structure from time-heterogeneous transition matrices. Whilst useful, these discussions focus only on the high-level formulation of each method, with little attention paid to model diagnostics, data structures, or implementation. Perhaps more importantly, \citet{skoglund2017} developed a few principles to which any term-structure model should adhere.
Firstly, a model should adequately capture the monotonicity of risk-ordering in that the poorest default risk can only improve or remain poor, and vice versa for the best default risk. Secondly, a model should be conditioned on the economy in that PD-estimates are to be sensitive to macroeconomic developments where applicable. Thirdly, a model should replicate observable experience with a relatively low degree of error.
We shall incorporate these principles into our work as far as possible.

\citet[\S3]{bellini2019} explored a range of modelling techniques, each of which is very briefly discussed and partly demonstrated in the R-programming language.
Their discussion starts with a framework wherein PD-estimates from the Basel context are re-used and adjusted with macroeconomic variables (MVs) using a \textit{generalised linear model} (GLM). A rudimentary survival probability is obtained from the predictions of this GLM, which is then used to derive lifetime PD-estimates; as we shall explore and critically review later in \autoref{sec:Basel_GLM}.
Thereafter, the author outlined and briefly demonstrated three methods from survival analysis that uses account-level data in continuous-time: 1) \textit{Kaplan-Meier} (KM) analysis; 2) \textit{Cox proportional hazards} (CPH) models; and 3) \textit{accelerated failure time} (AFT) models.
Other than survival analysis, three machine learning techniques (random forest, gradient boosting machine, and random survival forest) are also applied, having regressed account-level default indicators on a small list of variables, including MVs.
Finally, the author outlined transition-type models, and particularly emphasised an approach based on a Markov chain.
These approaches are reasonably illustrated and we commend the author for being one of the first to cover lifetime PD-modelling under IFRS 9. 
However, certain aspects can be expanded upon towards aligning with real-world applications. For example, the material on survival analysis -- including the description of data structures -- could certainly benefit from a more in-depth exploration of credit-relevant topics such as competing risks, left-truncation, and recurrent default events. Additionally, the underlying machine learning methods are introduced but were neither described, discussed, nor compared with one another. Finally, the model diagnostics, particularly for survival and machine learning models, could have been formulated in a more integrated, rigorous, and comprehensive manner. Our work aims to address these aspects by providing more detailed formulations and practical illustrations.

\citet{bank2021review} systematically considered 52 contributions to lifetime PD-modelling from 2013 up to 2020, having mixed them across wholesale and retail lending. They synthesised these contributions into three main classes of techniques: 1) survival analysis; 2) transition-type models (e.g., Markov chains); and 3) market-based models (e.g., Merton-type models). The former two classes differ mainly in that survival models assume a static state into which all loans are eventually absorbed (e.g., default), whereas transient states (e.g., default vs prepayment) can be handled by transition-type (or multistate) models. Survival models are further subdivided into their continuous-- and discrete-time varieties. Being relatively more well-known, the continuous-time variety include the non-parametric KM-estimator, the semi-parametric CPH-model, and two parametric models (the AFT-model and the proportional odds model). The authors first provided an overview of survival analysis, whereafter they briefly formulated and discussed each of these continuous-time techniques, including their associated merits and limitations.
The discussion then drifts towards the discrete-time variety, which is arguably the better variety in catering for credit data\footnote{The underlying data-generating mechanism of credit data is typically discrete in nature; i.e., interval-censored monthly observations.}. In essence, a discrete-time hazard/survival model directly embeds the baseline hazard $h_0(t)$ over event time $t$ as a series of input variables within a broader GLM-structure, as we shall review later in \autoref{sec:survival}.
\citet{bank2021review} have certainly provided a noteworthy synthesis of literature. However, and considering their paper's sizeable length, they were naturally constrained in further providing data-driven examples of each technique, or model diagnostics for that matter.

Even if somewhat outside of IFRS 9, \citet[\S 6]{baesens2016credit} further described and illustrated various \textit{discrete-time hazard} (DtH) models in modelling the lifetime PD using the SAS programming language. These models were mainly differentiated by the choice of link function within a GLM-setup: linear, logit, probit (the main model), and complementary log-log (cloglog). The authors provided both an overview of the required panel data, as well as some diagnostics for evaluating these models: \textit{Akaike's Information Criterion} (AIC), a likelihood-based pseudo coefficient of determination ($R^2$), and the \textit{area under the curve} (AUC) in summarising a \textit{receiver operating characteristic} (ROC) analysis; itself explained by \citet{fawcett2006introduction}.
While the choice of link function is stated to be arbitrary, the diagnostics differed admittedly across the link functions, with cloglog slightly underperforming both logit and probit. The latter two functions are purportedly more popular in practice, and we note that a special relationship exists with continuous-time survival models when using logit; see \citet{cox1972regression} and \citet{suresh2022survival}.
Regardless, the models of \citet[\S 6]{baesens2016credit} do not embed the baseline hazard $h_0(t)$ as an explicit input variable(s), and so their characterisation as survival models, as described by \citet{singer1993time} and \citet{suresh2022survival}, become somewhat questionable\footnote{Instead, the authors stated that their DtH-models "explain the default event during a certain time period". However, their formulation and interpretation of such models largely follow that of DtH-models, despite the difference in terminology.}.
Other than DtH-models, \citet[\S 7]{baesens2016credit} briefly discussed and illustrated a variety of continuous-time survival modelling techniques in estimating the lifetime PD. These techniques included: 1) two non-parametric methods (the KM-estimator and the actuarial life table method); 2) the semi-parametric CPH-model; and 3) the AFT-model. However, and as in the case of \citet{bank2021review}, the authors also recommended DtH-models over their continuous-time counterparts, given the latter's relatively greater degree of complexity whilst performing similarly.

\subsection{The use of survival analysis in lifetime PD-estimation}
\label{sec:background_surv}

Survival analysis has a long but still relatively sparse history of being used in modelling elements of credit risk. We refer the interested reader to \citet{singer1993time}, \citet{kleinbaum2012survival}, \citet{kartsonaki2016survival}, \citet{crowder2012credit}, \citet[\S 5]{thomas2017credit}, and \citet{schober2018survival} for a comprehensive discussion of survival analysis, which is more commonly used in the biostatistical literature than the credit domain. Survival analysis was first used in modelling the PD by \citet{narain1992credit}, whereafter \citet{banasik1999not} expanded thereon by using a CPH-model with input variables. \citet{stepanova2002survival} further investigated certain CPH-modelling practices in building a PD-model, as well as associated diagnostics (e.g., Cox-Snell and Schoenfeld residuals). Various other authors have demonstrated that including time-dependent variables, especially macroeconomic variables (MVs), can improve a CPH-based PD-model; see \citet{bellotti2009macro} and \citet{crook2010dynamic}. In particular, both \citet{bellotti2013} and \citet{bellotti2014stresstesting} used DtH-models with a logit link together with MVs, whilst embedding discretised time intervals using four different transforms towards incorporating the baseline hazard. They found that such models provided more accurate forecasts within a stress-testing setup, as evaluated using a few diagnostics such as the deviance statistic and Cox-Snell residuals. This result already agrees with both the second (macroeconomic sensitivity) and third (prediction accuracy) principles of \citet{skoglund2017} in producing a credible term-structure of default risk, which bodes well for using survival analysis.

Considering more recent literature on survival analysis in credit risk modelling, \citet{dirick2017time} examined a few subtypes thereof, including CPH-models with/without spline\footnote{In general, a spline function is defined as a smoothed composite of a series of basis functions, each of which is accompanied by an estimable coefficient; see \citet{perperoglou2019review} for an in-depth review.} functions on certain input variables, accelerated failure time models, and mixture cure models in modelling the lifetime PD. Unlike \citet{baesens2016credit}, their diagnostics included time-dependent ROC-analysis from \citet{heagerty2000}, which caters for right-censored observations when evaluating the discriminatory power of a survival model.
\citet{dirick2017time} also investigated the use of B-splines in capturing the nonlinearity between certain inputs and the hazard. They found that the CPH-model (with or without splines) generally outperformed the other techniques.
In building DtH-models with a cloglog link function,
\citet{djeundje2019dynamic} further explored B-splines by using them to estimate time-varying coefficients, including those of the baseline hazard $h_0(t)$. The premise is that the effect of a risk factor on the default hazard does not necessarily remain constant over the life of a loan; which is intuitively sensible, particularly so for longer-dated loan products such as mortgages. Their results showed that models with time-varying coefficients generally outperformed those without them.

Using US mortgage data, \citet{breeden2022multihorizon} proposed a variant of a DtH-model that is fitted to the lagged values of a delinquency variable in predicting the default hazard over different time horizons. This experimental setup resulted in 13 different DtH-based submodels, one for each lag, all of which formed one single "multi-horizon survival model". By building models using these lags, they argued that one circumvents the forecasting\footnote{This forecasting is typically necessary when using time-varying covariates within a prediction task.} of certain behavioural input variables.
They compared the predictions of this model to those of a few "matrix-type" models as a baseline, some of which are discussed later (i.e., roll rate models and transition-type models). They showed that their proposed model had the lowest forecast error and outperformed the other models over both shorter and longer time horizons. Whilst crucial to their experimental setup, the prospect of building (and maintaining) such a large of number of DtH-models is practically daunting and certainly invites model risk.
Lastly, \citet{botha2025recurrentEvents} investigated a few subtypes of CPH-models in dealing with recurrent defaults, i.e., the Andersen-Gill (AG) and the Prentice-Williams-Peterson (PWP) spell-time models. They found that the PWP-model performed similarly to a baseline model that deliberately ignored such recurrent defaults, whereas the AG-model underperformed expectations. Recurrent defaults are themselves a rarity and the effect of their inclusion into the modelling setup therefore depends on their prevalence, as corroborated by the authors. Nonetheless, our tutorial shall cater for recurrent defaults in the interest of completeness.

There are more complex extensions of the CPH-model within credit risk modelling, starting with the work of \citet{leow2014intensity}. The authors used CPH-models within a multistate setup (i.e., intensity\footnote{\citet{putter2007tutorial} described CPH-based intensity models as a generalisation of classical CPH-models within a multistate environment; also called competing risks.} models), where each transition type is modelled using a separate CPH-model with time-varying covariates, as estimated with retail credit card data. These CPH-based intensity models demonstrably produced predictions across any future time period and transition type, rather than merely the single-event default hazard of a standard CPH-model. In particular, the models are used in predicting the transition probability $p_{kl}(t, \boldsymbol{x}_{it}),i=1,\dots,n$ of moving from state $k$ to $l$ at time $t$ for loan $i$ with time-varying inputs $\boldsymbol{x}_{it}$. The authors found that while these models rendered fairly accurate predictions in general, the accuracy suffered at the loan-level.
\citet{djeundje2018intensity} extended this work by demonstrating the use of highly flexible B-splines in embedding the baseline intensities, even though doing so reportedly carries a high computational cost. They also showed how one might include random effects to account for unobserved heterogeneity, though their inclusion did not improve prediction accuracy.
Similarly, \citet{kelly2016good} modelled the PD and the probability of curing within a two-state framework, having used CPH-based intensity models with Irish residential mortgage data.

Another extension of CPH-models can be found within the framework of \textit{joint models}, which simultaneously model a longitudinal process and a time-to-event outcome. Hailing from biostatistics, joint models can model the association between repeated measurements of a longitudinal biomarker and patient survival, both of which share an unobservable underlying process. Such longitudinal outcomes arise from a so-called \textit{endogenous} variable, whose evolution is driven by some latent phenomenon (e.g., financial stress) that is closely related to the event of interest (e.g., default). Put differently, the latent process underlying the endogenous variable acts as both the data-generating mechanism for the longitudinal response and as a covariate governing event risk. The work of \citet{hu2018joint} and \citet{MedinaOlivares2023JointModels} have demonstrated that a joint model can outperform a CPH-model in terms of discrimination power and calibration quality, particularly within the context of behavioural credit scoring. A joint model therefore offers a robust framework for capturing dynamic (but latent) borrower behaviour, which can in turn improve prediction accuracy.
However, and despite these advantages, a joint model presents a few practical challenges. Firstly, it requires a complex model specification and its estimation is computationally intensive and relatively complicated. Secondly, a joint model cannot easily scale to large datasets, which are quite prevalent in credit risk modelling. Thirdly, it requires strong modelling assumptions, such as assuming that random effects are normally distributed, or assuming some pre-specified functional form for the association between the longitudinal and survival processes. Lastly, it has reduced transparency and interpretability that are otherwise crucial within regulatory environments (i.e., Basel-models), or those contexts that are subject to regular audit and governance requirements (i.e., IFRS 9 ECL-models). For these reasons, a joint model may be considered as overly complex when compared to a simpler alternative, such as a DtH-model.

\subsection{Matrix-type models and other approaches to lifetime PD-estimation}
\label{sec:background_other}

Loan-level models with a Markovian or transition-based structure have their own lineage in literature, which can comfortably accommodate both recurrent defaults and competing risks. Starting with the work of \citet{smith1995forecasting}, the authors built a nonstationary model for each transition type within a broader four-state transition framework: 1-Current and 2-Delinquent, and the absorbing states, 3-Written-off and 4-Settled. They considered two competing regression models within each cell, \textit{ordinary least squares} (OLS) and \textit{multinomial logistic regression} (MLR). In either case, loan-level state transitions were successfully predicted using a fairly rich and varied input space. However, it may be argued that not every cell within such a framework merits a separate model, especially if certain transition probabilities can rather be fixed given a lack of data. 
In fact, \citet{grimshaw2011markov} extended this work in exactly this way. They developed stratified binary logistic regression models within only the most crucial of cells of the transition matrix, though did not compare their model with a fully-fledged MLR-model.
\citet{Arundina2015sukuk} compared a four-state MLR-model against a neural network in predicting credit rating migrations (including default), having used data on Sukuk corporate bonds. Although they found that both techniques performed similarly, they used a rather rudimentary accuracy measure, which is notoriously influenced by class imbalance; all of which may affect their findings.
\citet{adha2018multinomial} favourably compared a three-state MLR-model against a parametric spline regression model, which was used with Indonesian credit card data in predicting the hazard of either default or attrition.
Using residential mortgage data, \citet{botha2025multistate} compared an MLR-model against a stationary Markov chain and a set of beta regression models, as estimated within a four-state transition-based framework. Their results showed that the MLR-model outperformed the other techniques on every metric, which included the average discrepancy between the empirical and expected term-structures of default risk.

Other lifetime PD-modelling methods covered by \citet{brunel2016lifetime} and \citet{bank2021review} include an \textit{age-period-cohort} (APC) model, which is a portfolio-level analysis of specific cohorts (or `vintages') called vintage models. In particular, the default rate is decomposed into three estimable functions: 1) the lifecycle function $F(\tau)$ that measures how default risk changes over loan age $\tau$; 2) the vintage function $G(v)$ that evaluates credit quality as a function of the origination date $v$; and 3) the environment function $H(t)$ that models exogenous macroeconomic conditions that affect all cohorts over calendar time $t$. Classical APC-models typically produce portfolio-level output, which implies assigning these predictions to a possibly heterogeneous array of 
individual loans. Doing so can degrade the loan-level prediction accuracy of APC-models, unless one pursues extensive segmentation procedures; itself not without its challenges.
That said, \citet{bank2021review} reviewed a loan-level variant of APC-models that can incorporate various input variables. Together with \citet{breeden2022multihorizon}, the authors claim that these models (essentially DtH-models) perform more robustly out-of-sample than competing models, though their estimation is certainly more complex and intricate.
Another modelling method is based on adjusting the output from standard behavioural credit scoring models, as reviewed by \citet{brunel2016lifetime}.
However, these scores can only predict 1-year PDs and are typically static in nature, which degrade their feasibility within longer-term and more dynamic setups such as under IFRS 9.

\citet{brunel2016lifetime} and \citet{bank2021review} also discussed "matrix models", of which the simplest kind is a roll rate model. This model is essentially a statistical tabulation of observed transitions amongst delinquency-based states over some period. However, this approach may yield statistically noisy estimates of a transition probability, particularly given the paucity of data for default-events. Moreover, this portfolio-level approach has limited predictive power in that it struggles to risk-sensitise its output to the characteristics of the individual borrower. Another assumption of this approach is that of time-homogeneity, such that its output is valid only when conditions and lending criteria remain constant over time; an unrealistic assumption in most cases.
Other than roll rate models, the authors reviewed rating-based matrix models that focus on the intensity $\lambda_{kl}$ by which borrowers migrate from state $k$ to $l$, instead of the likelihood of the event. However, classical intensity models typically consider only the time spent in $k$ (which is exponentially distributed), and cannot easily incorporate other inputs such as macroeconomic variables.
It is certainly true that intensity models can produce a richer description of the default process than survival models, and even generate non-zero PD-estimates in the absence of any defaults. However, this benefit is afforded at the cost of additional restrictions (e.g., the Markov assumption), greater complexity in the mathematical formalism, as well as a greater risk of overfitting the data in retail portfolios. Overall, it is perhaps a modelling technique that is best reserved for data-poor environments such as wholesale lending.

Ultimately, we think that the literature on using survival analysis is particularly convincing in using it as a main method for lifetime PD-estimation under IFRS 9. While other methods certainly exist, they each bring about a certain challenge. Either they are relatively complex and time-intensive to build or maintain, or they are not sufficiently complex in rendering credible PD-estimates. So far, the studies on survival analysis augur well for positioning this class of techniques as a middle-ground of sorts, which can produce credible lifetime PD-estimates without undue complexity or effort. It is for this reason that we shall focus our efforts on using survival analysis, particularly on building a DtH-model with time-fixed and time-varying covariates (such as MVs), replete with model diagnostics.

%% file: 3-SurvivalModelling.tex
\section{Estimating lifetime PDs using survival analysis in discrete time}
\label{sec:survival}

Basic concepts, related notation, data structures, a censoring study, and associated event time distributions are explored in \autoref{sec:survival_concepts}. These aspects are all formulated and illustrated within the specific context of credit risk modelling. 
Thereafter, an appropriate resampling scheme is discussed in \autoref{sec:survival_resampling} towards preparing the data for survival modelling. This discussion includes the formulation and demonstration of a diagnostic measure (the \textit{resolution rate}) for evaluating the representativeness of a resampling scheme in guarding against sampling bias.
The fundamentals of discrete-time survival analysis are surveyed in \autoref{sec:survival_fundamentals}, followed by an illustration of the empirical term-structure of default risk.
In \autoref{sec:survival_discTimeCoxModel}, we formulate and implement a \textit{discrete-time hazard} (DtH) model within the context of recurrent default events and class imbalance. Applicable diagnostics are illustrated in \autoref{sec:survival_diagnostics} towards evaluating the model amidst right-censoring, which include \textit{time-dependent }ROC-analysis (tROC), and the \textit{time-dependent Brier score} (tBS). Other calibration-focused diagnostics include empirical vs expected comparisons of both term-structures and of 12-month default rates.
Each of these subsections may be viewed as particular steps in building a DtH-model for estimating the term-structure of default risk. Lastly, this tutorial leverages the real-world credit histories of 90,000 randomly subsampled mortgage accounts from the South African market, spanning the years 2007-2022, which are consumed by the R-codebase of \citet{botha2025termStructureSourcecode}.

\input{3.1-SurvivalData}
\input{3.2-ResamplingScheme}
\input{3.3-Fundamentals}
\input{3.4-DiscTimeCoxModel}

\input{3.5-Diagnostics}

\input{3.6-Considerations}

%% file: 3.1-SurvivalData.tex
\subsection{Towards structuring credit survival data: Basic concepts \& notation}
\label{sec:survival_concepts}

\begin{figure}[ht!]
    \centering
    \includegraphics[width=1\linewidth, height=0.38\textheight]{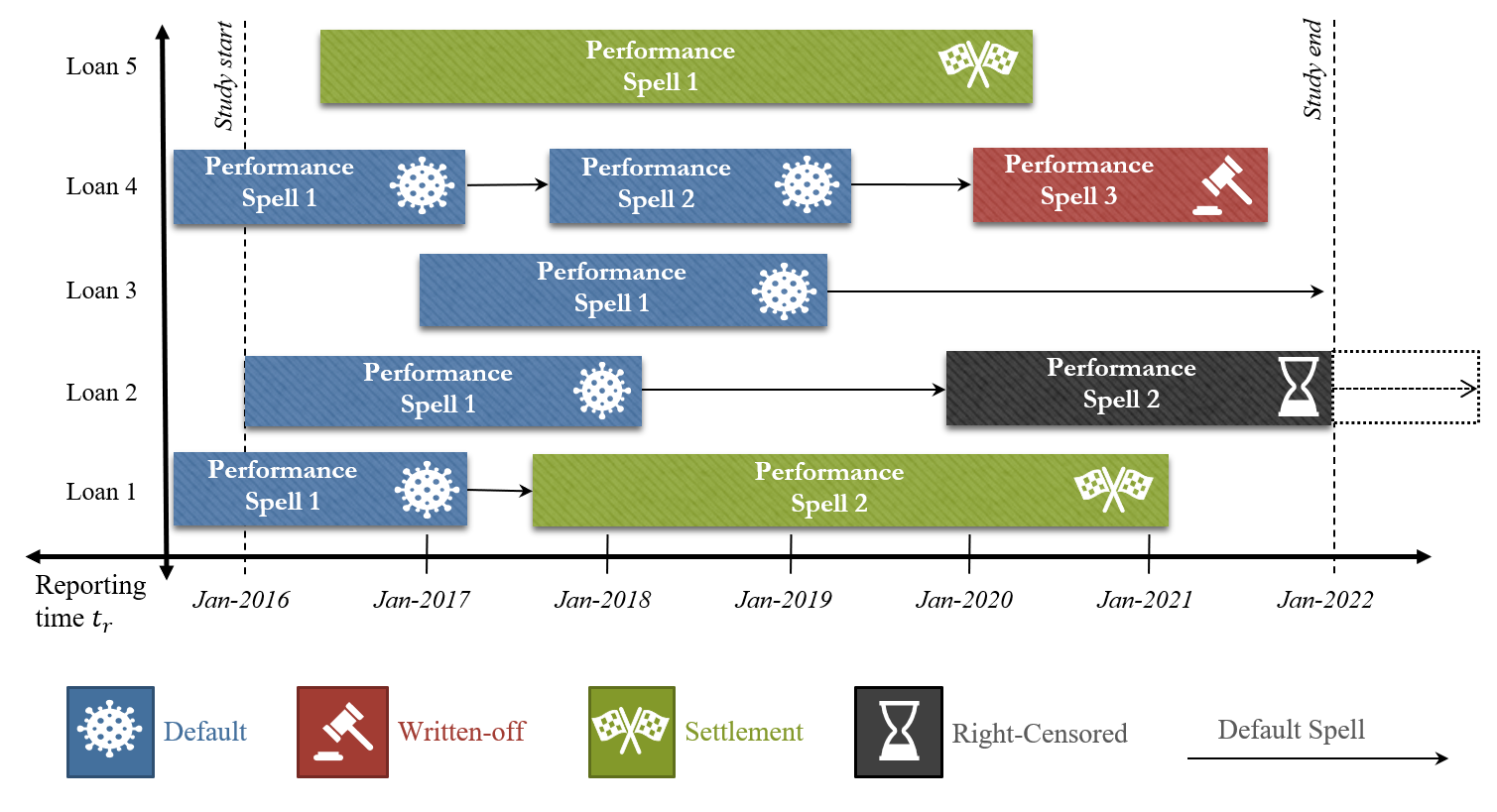}
    \caption{Demonstrating the resolution types of performing spells over time for a few hypothetical loans. From \citet{botha2025recurrentEvents}.}
    \label{fig:PerfSpells}
\end{figure}

In formalising the use of survival analysis within the credit domain, we shall start with some notation that represents the data structure and its salient features, as inspired by \citet{botha2025recurrentEvents}. 
A loan may experience multiple types of events during its lifetime and its experience can be bifurcated into either performance spells or default\footnote{Default spells are the purview of LGD-modelling, and we shall therefore not explore its formulation given our focus on lifetime PD-modelling. Nonetheless, their formulation is very similar.} spells. A \textit{spell} represents a multi-period time span during which time a bank monitors the repayment performance of a loan, which ends in some resolution outcome. Naturally, a performance spell starts at entry time $\tau_e$ from the moment of account origination, and ends at a resolution time $\tau_r > \tau_e$, usually coinciding with the default event. That said, a defaulted loan can cure and become subject to default risk once more, all of which implies a `multi-spell' setup (or recurrent survival analysis) in tracking the loan over its lifetime; see \citet{willett1995} and \citet[\S1.1]{jenkins2005survival}.
Furthermore, `default' is not the only state into which a loan may resolve, and other outcomes exist alongside it, such as write-off and early settlement. These \textit{competing risks} preclude the default event from occurring and their existence has bearing on the overall modelling of default risk. We illustrate these ideas on performing spells in \autoref{fig:PerfSpells} for a few hypothetical loans across various competing possibilities in resolving each loan.

More formally, consider a portfolio of $N_p$ loans, wherein any loan $i=1,\dots,N_p$ may have $j=1,\dots,n_i \geq 1$ number of performing spells. The subject-spell construct $(i,j)$ uniquely denotes the portion of the overall loan history that represents a single performance spell, accompanied by a resolution outcome.
Some spells may lack such a resolution outcome and we denote such \textit{right-censored} cases, as discussed later, with $c_{ij}\in\{0,1\}$ in that $c_{ij}=1$ for a right-censored spell $(i,j)$, and $c_{ij}=0$ otherwise.
The outcomes into which $(i,j)$ may resolve can be coalesced into a single nominal variable $\mathcal{R}_{ij}$, which is encoded as
\begin{align} \label{eq:spellResolution_Types}
    \mathcal{R}_{ij} = 
    \begin{cases}
        1: \text{Default} \quad & \text{if} \ c_{ij}=0 \ \text{and default-criteria applies} \\
        2: \text{Settled} \quad & \text{if} \ c_{ij}=0 \ \text{and settlement-criteria applies} \\
        3: \text{Write-off/Other} \quad & \text{if} \ c_{ij}=0 \ \text{and write-off (or other) criteria applies} \\        
        4: \text{Censored} \quad & \text{if} \ c_{ij} = 1
    \end{cases} \, .
\end{align}

Regarding timing, a performing spell $(i,j)$ is observable from its entry time $\tau_e(i,j)\geq 0$, and observation continues either up to the spell resolution time $\tau_r(i,j)$ for $c_{ij}=0$, or up to the censoring time $C_{ij}<\tau_r(i,j)$ for $c_{ij}=1$. Taking the minimum between $\tau_r(i,j)$ and $C_{ij}$ would give the overall spell stopping time $\tau_s(i,j)$. Time is itself measured discretely during a spell using an integer-valued counter variable, which is called the spell period $t_{ij}=\tau_e(i,j),\dots,\tau_s(i,j)$ for spell $j$ of loan $i$.
By denoting time with an explicit start and stopping point, and using the lexicon of survival analysis, the data is structured in the so-called \textit{counting process} style, as discussed by \citet[pp.~20-23]{kleinbaum2012survival}.
The overall age of $(i,j)$ is then defined as the observable failure (or default) time $T_{ij}$, or follow-up time, expressed using starting/stopping times as
\begin{equation} \label{eq:timeSpent_performing}
    T_{ij} = \min\Big(\tau_r(i,j), C_{ij}\Big) - \tau_e(i,j) = \tau_s(i,j) - \tau_e(i,j) \, .
\end{equation}
Put differently, spell $(i,j)$ has failure/event time $T_{ij}\in \mathbb{Z}_{\geq1}$ at which either the default event (or another competing risk) occurred, or the subject was right-censored. We shall henceforth drop the $(i,j)$-part from the notation of certain quantities $\left\{\tau_e, \tau_s, \tau_r \right\}$ in the interest of simplicity, though the dependence on a particular spell remains implied. 
We further differentiate the spell period $t_{ij}$ from the overall loan period $t_i$, which simply tracks the calendar time of the loan's lifetime (e.g., "time on book"), regardless of performing or default spells. This spell period is of vital importance since it effectively represents the time spent in $(i,j)$ at each point of its duration; a quantity that will surface again later.
Finally, the event history indicator $e_{ijt}$ flags whether or not the main (default) event occurred at a specific spell period $t_{ij}$, where zero-values indicate either right-censoring or competing risks. \citet{singer1993time} described $e_{ijt}$ as a "chronology of event indicators" since it produces either a vector $(0,0,\dots,1)$ for a defaulted subject-spell, or $(0,0,\dots,0)$ for a censored subject-spell; formalised later in \autoref{sec:survival_fundamentals}. Lastly, we note that encoding competing risks as right-censored cases in this way is known as the \textit{latent risks} approach, as discussed by \citet{putter2007tutorial}.

The main thrust of survival analysis is to investigate the random variable $T\geq 0$ that represents the latent lifetimes of performing spells until reaching some well-defined failure point (e.g., default). However, this aim can never be truly realised by simply using the observed spell ages from \autoref{eq:timeSpent_performing}, because of the confounding possibilities of \textit{censoring} and \textit{truncation}; as explained by \citet[\S1.3]{jenkins2005survival}. While censoring is broadly interpreted as the \textit{absence} of observing an outcome during some time span, a few types of censoring exist; see \citet[\S 1]{kleinbaum2012survival} and \citet{schober2018survival}. The most common type is that of \textit{right-censoring} where the true (but unobservable) end of an incomplete spell $(i,j)$ is `right-most' of the study-end, as the black-shaded spell illustrates in \autoref{fig:PerfSpells}. While a completed spell ending in default would suggest $T=T_{ij}$, a right-censored spell suggests $T>T_{ij}$, which certainly complicates the estimation of $T$.
Another nuisance is the prevalence of \textit{left-truncated} spells within a typically-sampled credit dataset. i.e., a spell whose starting point predates that of the overall sampling window; see \citet[\S 1.2.1]{jenkins2005survival} and \citet[pp.~132-134]{kleinbaum2012survival}. In practice, a loan portfolio may have extensive left-truncation, especially so for longer-dated products such as residential mortgages. While the truncated histories of such spells may exist theoretically, it is often impractical or infeasible to attempt their sampling beyond a certain historical point, e.g., due to system constraints. Nonetheless, we encode the presence of left-truncation by adjusting the starting time of the first affected spell. In particular, $\tau_e(i,j)$ is expediently set to the inferred loan age at the start time of observation, where loan age is itself calculated as the difference between the current date and the origination date.

The ideas presented thus far culminate in the longitudinal dataset $\mathcal{D}=\left\{i,t_i,j, t_{ij}, \tau_e, \tau_s, \mathcal{R}_{ij}, T_{ij}, e_{ijt} \right\}$, as illustrated in \autoref{tab:dataStructure_perfSpells_PWP} for a few hypothetical performing loans. Note that these loans are not the same ones as in \autoref{fig:PerfSpells}.
Each subject-spell $(i,j)$ in the table has multiple observations, where each row is uniquely identified by the composite key $(i,j,t_{ij})$. For example, Loan 3 ($i=3$) had two performing spells; the first spell ended in default while the second spell ended in settlement, after spending four and three months in the performing state respectively. Loan 4 ($i=4$) had a delayed entry (i.e., it was left-truncated) at month $t_i=5$ for spell $j=1$, which is why its entry time is adjusted to $\tau_e=4$, assuming that it was still in performing prior. It then defaulted $T_{ij}=5$ months later at $t_i=9$, followed by two successive performing spells; the last of which became right-censored at time $t_i=41$.
The fact that the spell period resets upon entering each successive spell in $\mathcal{D}$ is a feature of a particular type of survival modelling that can handle recurrent events, i.e., the \textit{Prentice-Williams-Peterson} (PWP) technique; as will be discussed later. 
In short, the ordering of successive spells becomes important, as does the way in which time is encoded within $\mathcal{D}$.
In this work, we shall similarly contend with the issue of recurrent default events, and therefore opt for the PWP-technique.

\begin{longtable}[ht!]{p{1cm} p{1.1cm} p{1.6cm} p{1.3cm} p{1.2cm} p{1.2cm} p{1.9cm} p{1.3cm} p{1.3cm}}
\caption{Illustrating the structure of the raw panel/longitudinal dataset $\mathcal{D}$ and its performing spells. The alternating grey-shaded rows indicate loan-level history, while the alternating colour-shaded cells signify the performing spell-level histories respective to each loan; the remaining unshaded cells denote period-level information. From \citet{botha2025recurrentEvents}.} \label{tab:dataStructure_perfSpells_PWP} \\
\toprule
\textbf{Loan} $i$ & \textbf{Period} $t_i$ & \textbf{Spell number} $j$ & \textbf{Spell period} $t_{ij}$ & \textbf{Entry time} $\tau_e$ & \textbf{Stop time} $\tau_s$ & \textbf{Resolution type} $\mathcal{R}_{ij}$ & \textbf{Spell age} $T_{ij}$ & \textbf{Event} $e_{ijt}$ \\ 
\midrule
\endfirsthead
\caption[]{(continued)} \\
\toprule
\textbf{Loan} $i$ & \textbf{Period} $t_i$ & \textbf{Spell number} $j$ & \textbf{Spell period} $t_{ij}$ & \textbf{Entry time} $\tau_e$ & \textbf{Stop time} $\tau_s$ & \textbf{Resolution type} $\mathcal{R}_{ij}$ & \textbf{Spell age} $T_{ij}$ & \textbf{Event} $e_{ijt}$ \\ 
\midrule
\endhead
\midrule \multicolumn{8}{r}{\textit{Continued on next page}} \\
\endfoot
\bottomrule
\endlastfoot
\cellcolor[HTML]{EFEFEF}1 & 1 & \cellcolor[HTML]{ECF4FF}1 & 1 & \cellcolor[HTML]{ECF4FF}0 & \cellcolor[HTML]{ECF4FF}4 & \cellcolor[HTML]{ECF4FF}1: Defaulted & \cellcolor[HTML]{ECF4FF}4 & 0 \\
\cellcolor[HTML]{EFEFEF}1 & 2 & \cellcolor[HTML]{ECF4FF}1 & 2 & \cellcolor[HTML]{ECF4FF}0 & \cellcolor[HTML]{ECF4FF}4 & \cellcolor[HTML]{ECF4FF}1: Defaulted & \cellcolor[HTML]{ECF4FF}4  & 0\\
\cellcolor[HTML]{EFEFEF}1 & 3 & \cellcolor[HTML]{ECF4FF}1 & 3 & \cellcolor[HTML]{ECF4FF}0 & \cellcolor[HTML]{ECF4FF}4 & \cellcolor[HTML]{ECF4FF}1: Defaulted & \cellcolor[HTML]{ECF4FF}4  & 0\\
\cellcolor[HTML]{EFEFEF}1 & 4 & \cellcolor[HTML]{ECF4FF}1 & 4 & \cellcolor[HTML]{ECF4FF}0 & \cellcolor[HTML]{ECF4FF}4 & \cellcolor[HTML]{ECF4FF}1: Defaulted & \cellcolor[HTML]{ECF4FF}4 & 1\\
\cellcolor[HTML]{C0C0C0}2 & 1 & \cellcolor[HTML]{C0DAFE}1 & 1 & \cellcolor[HTML]{C0DAFE}0 & \cellcolor[HTML]{C0DAFE}3 & \cellcolor[HTML]{C0DAFE}4: Censored & \cellcolor[HTML]{C0DAFE}3  & 0\\
\cellcolor[HTML]{C0C0C0}2 & 2 & \cellcolor[HTML]{C0DAFE}1 & 2 & \cellcolor[HTML]{C0DAFE}0 & \cellcolor[HTML]{C0DAFE}3 & \cellcolor[HTML]{C0DAFE}4: Censored & \cellcolor[HTML]{C0DAFE}3  & 0\\
\cellcolor[HTML]{C0C0C0}2 & 3 & \cellcolor[HTML]{C0DAFE}1 & 3 & \cellcolor[HTML]{C0DAFE}0 & \cellcolor[HTML]{C0DAFE}3 & \cellcolor[HTML]{C0DAFE}4: Censored & \cellcolor[HTML]{C0DAFE}3 & 0\\
\cellcolor[HTML]{EFEFEF}3 & 1 & \cellcolor[HTML]{ECF4FF}1 & 1 & \cellcolor[HTML]{ECF4FF}0 & \cellcolor[HTML]{ECF4FF}4 & \cellcolor[HTML]{ECF4FF}1: Defaulted & \cellcolor[HTML]{ECF4FF}4  & 0\\
\cellcolor[HTML]{EFEFEF}3 & 2 & \cellcolor[HTML]{ECF4FF}1 & 2 & \cellcolor[HTML]{ECF4FF}0 & \cellcolor[HTML]{ECF4FF}4 & \cellcolor[HTML]{ECF4FF}1: Defaulted & \cellcolor[HTML]{ECF4FF}4 & 0 \\
\cellcolor[HTML]{EFEFEF}3 & 3 & \cellcolor[HTML]{ECF4FF}1 & 3 & \cellcolor[HTML]{ECF4FF}0 & \cellcolor[HTML]{ECF4FF}4 & \cellcolor[HTML]{ECF4FF}1: Defaulted & \cellcolor[HTML]{ECF4FF}4 & 0 \\
\cellcolor[HTML]{EFEFEF}3 & 4 & \cellcolor[HTML]{ECF4FF}1 & 4 & \cellcolor[HTML]{ECF4FF}0 & \cellcolor[HTML]{ECF4FF}4 & \cellcolor[HTML]{ECF4FF}1: Defaulted & \cellcolor[HTML]{ECF4FF}4 & 1 \\
\cellcolor[HTML]{EFEFEF}3 & 11 & \cellcolor[HTML]{E6FFE6}2 & 1 & \cellcolor[HTML]{E6FFE6}0 & \cellcolor[HTML]{E6FFE6}3 & \cellcolor[HTML]{E6FFE6}2: Settled & \cellcolor[HTML]{E6FFE6}3 & 0 \\
\cellcolor[HTML]{EFEFEF}3 & 12 & \cellcolor[HTML]{E6FFE6}2 & 2 & \cellcolor[HTML]{E6FFE6}0 & \cellcolor[HTML]{E6FFE6}3 & \cellcolor[HTML]{E6FFE6}2: Settled & \cellcolor[HTML]{E6FFE6}3 & 0 \\
\cellcolor[HTML]{EFEFEF}3 & 13 & \cellcolor[HTML]{E6FFE6}2 & 3 & \cellcolor[HTML]{E6FFE6}0 & \cellcolor[HTML]{E6FFE6}3 & \cellcolor[HTML]{E6FFE6}2: Settled & \cellcolor[HTML]{E6FFE6}3 & 0 \\
\cellcolor[HTML]{C0C0C0}4 & 5 & \cellcolor[HTML]{C0DAFE}1 & 5 & \cellcolor[HTML]{C0DAFE}4 & \cellcolor[HTML]{C0DAFE}9 & \cellcolor[HTML]{C0DAFE}1: Defaulted & \cellcolor[HTML]{C0DAFE}5  & 0\\
\cellcolor[HTML]{C0C0C0}4 & 6 & \cellcolor[HTML]{C0DAFE}1 & 6 & \cellcolor[HTML]{C0DAFE}4 & \cellcolor[HTML]{C0DAFE}9 & \cellcolor[HTML]{C0DAFE}1: Defaulted & \cellcolor[HTML]{C0DAFE}5 & 0\\
\cellcolor[HTML]{C0C0C0}4 & 7 & \cellcolor[HTML]{C0DAFE}1 & 7 & \cellcolor[HTML]{C0DAFE}4 & \cellcolor[HTML]{C0DAFE}9 & \cellcolor[HTML]{C0DAFE}1: Defaulted & \cellcolor[HTML]{C0DAFE}5 & 0 \\
\cellcolor[HTML]{C0C0C0}4 & 8 & \cellcolor[HTML]{C0DAFE}1 & 8 & \cellcolor[HTML]{C0DAFE}4 & \cellcolor[HTML]{C0DAFE}9 & \cellcolor[HTML]{C0DAFE}1: Defaulted & \cellcolor[HTML]{C0DAFE}5 & 0 \\
\cellcolor[HTML]{C0C0C0}4 & 9 & \cellcolor[HTML]{C0DAFE}1 & 9 & \cellcolor[HTML]{C0DAFE}4 & \cellcolor[HTML]{C0DAFE}9 & \cellcolor[HTML]{C0DAFE}1: Defaulted & \cellcolor[HTML]{C0DAFE}5 & 1 \\
\cellcolor[HTML]{C0C0C0}4 & 20 & \cellcolor[HTML]{B5FFB5}2 & 1 & \cellcolor[HTML]{B5FFB5}0 & \cellcolor[HTML]{B5FFB5}4 & \cellcolor[HTML]{B5FFB5}1: Defaulted & \cellcolor[HTML]{B5FFB5}4 & 0 \\
\cellcolor[HTML]{C0C0C0}4 & 21 & \cellcolor[HTML]{B5FFB5}2 & 2 & \cellcolor[HTML]{B5FFB5}0 & \cellcolor[HTML]{B5FFB5}4 & \cellcolor[HTML]{B5FFB5}1: Defaulted & \cellcolor[HTML]{B5FFB5}4 & 0 \\
\cellcolor[HTML]{C0C0C0}4 & 22 & \cellcolor[HTML]{B5FFB5}2 & 3 & \cellcolor[HTML]{B5FFB5}0 & \cellcolor[HTML]{B5FFB5}4 & \cellcolor[HTML]{B5FFB5}1: Defaulted & \cellcolor[HTML]{B5FFB5}4 & 0 \\
\cellcolor[HTML]{C0C0C0}4 & 23 & \cellcolor[HTML]{B5FFB5}2 & 4 & \cellcolor[HTML]{B5FFB5}0 & \cellcolor[HTML]{B5FFB5}4 & \cellcolor[HTML]{B5FFB5}1: Defaulted & \cellcolor[HTML]{B5FFB5}4 & 1 \\
\cellcolor[HTML]{C0C0C0}4 & 40 & \cellcolor[HTML]{FFE1BD}3 & 1 & \cellcolor[HTML]{FFE1BD}0 & \cellcolor[HTML]{FFE1BD}2 & \cellcolor[HTML]{FFE1BD}4: Censored & \cellcolor[HTML]{FFE1BD}2 & 0 \\
\cellcolor[HTML]{C0C0C0}4 & 41 & \cellcolor[HTML]{FFE1BD}3 & 2 & \cellcolor[HTML]{FFE1BD}0 & \cellcolor[HTML]{FFE1BD}2 & \cellcolor[HTML]{FFE1BD}4: Censored & \cellcolor[HTML]{FFE1BD}2 & 0 \\
\end{longtable}

\begin{figure}[ht!]
    \centering
    \includegraphics[width=0.8\linewidth,height=0.45\textheight]{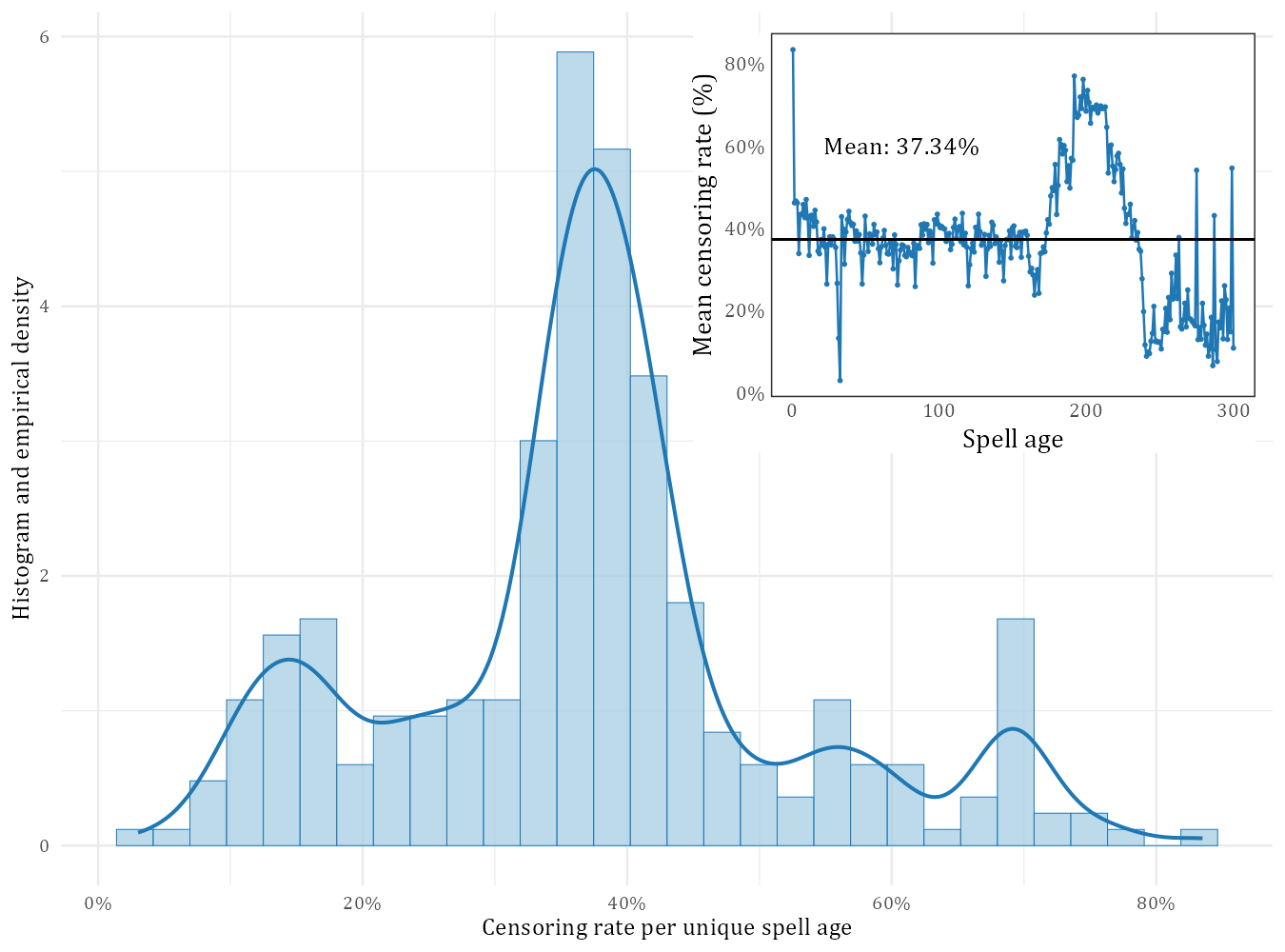}
    \caption{Histogram of the calculated censoring degrees across spell ages $T_{ij}\leq 300$, having used residential mortgage data. The inset graph shows the average censoring rate per unique spell age, whereas the main graph summarises its statistical distribution.}
    \label{fig:Hist_CensoringDegree}
\end{figure}

The prevalence of right-censoring is perhaps the most convincing argument for adopting a survival-type technique in modelling the term-structure of default risk. As an alluring option, one might be tempted to exclude such right-censored cases from the modelling dataset. However, doing so not only amounts to an inefficient use of data, but also poses significant model risk, particularly when the censored population is relatively large in size.
In fact, \citet{watt1996censored} examined the effect of censoring on survival probability estimates by comparing crude estimates (that ignore censoring) with those from a Kaplan-Meier study (which incorporates censoring). This crude estimator is simply the number of subjects that survive beyond a certain time point, divided by the volume of at-risk accounts at the same time. Overall, the authors found consistent underestimation within this crude estimator, relative to the true survival probability that is obtained from the Kaplan-Meier estimator. The level of such underestimation also grew as the degree of censoring increased within the sample.
As such, prudence dictates that one should investigate the degree of censoring within a given dataset, especially in motivating the use of a survival-type modelling setup for the PD term-structure. We provide an example of such a censoring study in \autoref{fig:Hist_CensoringDegree}, having used residential mortgage data that was reasonably capped to a maximum of 300-month long spells. Evidently, the censoring rates across spell ages are unexpectedly high, with a mean censoring rate of about 37\% across all unique spell ages. However, the censoring rate is particularly high at both early and later spell ages, which suggests that one should be cautious when interpreting this mean on its own. In accordance with \citet{watt1996censored}, we suggest having (and imposing) a mean censoring level of at least 20\% when motivating a survival-type study, thereby contending with the resulting right-censored cases and their high prevalence.

\begin{figure}[ht!]
    \centering
    \includegraphics[width=0.8\linewidth,height=0.45\textheight]{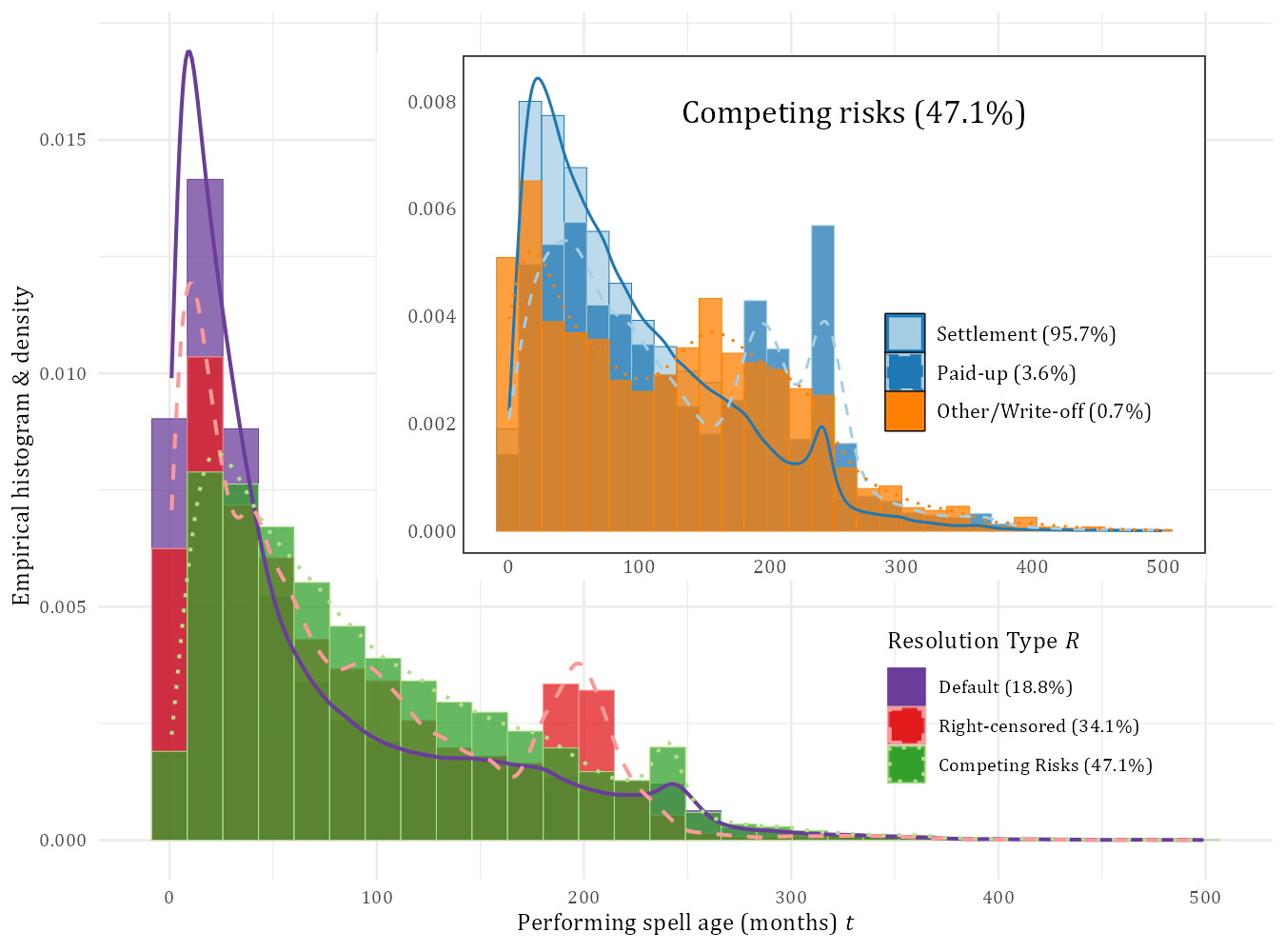}
    \caption{Histograms of failure times (or spell ages) for $T_{ij}\leq 500$ by resolution type, having used residential mortgage data. Empirical density estimates are overlaid, while the inset graph shows histograms and density estimates for all competing risks that preclude the main event (default) from occurring.}
    \label{fig:Hist_PerfSpellAges}
\end{figure}

\citet[p.~42]{kleinbaum2012survival} and \citet{schober2018survival} explained that censoring should be \textit{non-informative} in that it does not influence the occurrence or timing of the main default event. Having non-informative censoring implies independence between the distributions of the censoring times $C_{ij}$ and the resolution times $\tau_r$. In this regard, we believe it sufficient to graph and compare the empirical histograms and densities of the various types of failure times, or loan ages $T_{ij}, i=1,\dots,N_p,j=1,\dots,n_i$ across resolution types $\mathcal{R}_{ij}$. Such a comparison is provided in \autoref{fig:Hist_PerfSpellAges} using the same data, and shows that the $C_{ij}$-distribution (in red) differs materially from the others; itself suggesting that censoring is indeed non-informative. Moreover, \autoref{fig:Hist_PerfSpellAges} is particularly useful as a diagnostic tool in understanding the effect and prevalence of competing risk-events, whose occurrence precludes the main default event. For example, default occurred only in about 19\% of spells, whilst 47\% thereof suffered a competing risk, most of which (96\%) were early settlements. The relatively high prevalence of these competing risks suggests that they ought to be incorporated in some fashion into the eventual modelling strategy, lest it provide biased results.

%% file: 3.2-ResamplingScheme.tex
\subsection{Resampling schemes and a diagnostic measure for testing sampling representativeness}
\label{sec:survival_resampling}

The prediction accuracy of any eventual survival model can be evaluated by estimating the test error on a specifically-designated validation dataset. Such a set is however rarely available, which is why resampling methods are crucial in creating it from the raw dataset $\mathcal{D}$ from \autoref{tab:dataStructure_perfSpells_PWP}. One particularly common group of resampling methods is that of \textit{simple cross-validation}, which is based on randomly splitting $\mathcal{D}$ into two non-overlapping sets, denoted respectively as $\mathcal{D}_T$ (for model training) and $\mathcal{D}_V$ (for model validation). As discussed by \citet[pp.~32-33]{bishop2006pattern} and \citet[pp.~176-178]{james2013introduction}, observations $i$ from $\mathcal{D}$ are randomly selected into $\mathcal{D}_T$ such that a chosen sampling fraction $s_f=|\mathrm{D}_T|/|\mathcal{D}|$ is maintained. In principle, one would like to reserve as much data as possible into $\mathcal{D}_T$ towards training the model optimally. However, too small a $\mathcal{D}_V$-set will lead to noisy and ultimately unreliable estimates of prediction accuracy and/or discriminatory power. To balance these two competing interests against each other is an imprecise art, though common practice suggests that $s_f\in[60\%, 80\%]$ when creating $\mathcal{D}_T$; see \citet[pp.~63-65,~127-128]{siddiqi2005credit}. 
In this section, we shall demonstrate a re-usable diagnostic measure by which resampled datasets and their representativeness can be tested over time.

In survival analysis, it is preferable to retain the entire spell history across all of its periods $t_{ij}$ when resampling from $\mathcal{D}$, lest the subsequent survival estimates become compromised due to missing spell periods. Row-observations in $\mathcal{D}$ therefore need to be \textit{clustered} around a common characteristic before sampling randomly amongst the resulting clusters. This sampling design is usually called \textit{simple random clustered sampling} and implies that all monthly observations in $\mathcal{D}$ are first grouped by the loan ID $i$, thereby forming $N_p$ clusters. Thereafter, we reserve a proportion ($s_f=70\%$) thereof for $\mathcal{D}_T$, whereupon the under lying credit histories of these randomly chosen clusters are sampled into $\mathcal{D}_T$, whilst relegating the rest into $\mathcal{D}_V$.  This sampling technique is also briefly discussed and illustrated by \citet[\S 6]{baesens2016credit} within a similar context.

In principle, the sets $\left\{ \mathcal{D}_T, \mathcal{D}_V \right\}$ should not exhibit undue sampling bias over time. In measuring such bias, consider that the resolution type $\mathcal{R}_{ij},i=1,\dots,N_p,j=1,\dots,n_i$ are spell-level realisations from a nominal-valued random variable $\mathcal{R}$. These realisations can be aggregated to the portfolio-level by first partitioning the panel dataset $\mathcal{D}$ into a series of non-overlapping monthly spell cohorts $\mathcal{D}(t')$ over reporting time $t'=t'_1,\dots,t_l',\dots,t'_n$. Consider $Y_\psi\in\{0,1\}$ as a Bernoulli random variable for a specific event type $\psi\in\mathcal{R}$. A series of such random variables exists over time $t'$ for each $\psi$, denoted as $Y_\psi\left(t'_1\right),\dots,,Y_\psi\left(t'_l\right)\dots,Y_\psi\left(t'_n\right)$. Let $r_\psi\left(t'_l,\mathcal{D}\right)$ denote the \textit{resolution rate} at which the modelled phenomenon resolves at any $t'_l$ into a specified type $\psi$ within a given dataset $\mathcal{D}$. This $r_\psi\left(t'_l,\mathcal{D}\right)$-quantity estimates the probability $\mathbb{P}\left(Y_\psi(t'_l) = 1 \right)$ within $\mathcal{D}$, and is intuitively calculated as the proportion of 1's in $Y_\psi$ of type $\psi$ at a particular time $t'_l$. If $n_{t'}$ denotes the size of the partitioned set $\mathcal{D}(t')$, then we formally define $r_\psi\left(t',\mathcal{D}\right)$ of type $\psi$ at each $t'$ as
\begin{equation} \label{eq:ResolRate}
    r_\psi(t',\mathcal{D}) = \frac{1}{n_{t'}} \sum_{(i,j) \, \in \, \mathcal{D}(t')} \mathbb{I}(\mathcal{R}_{ij}= \psi) \quad \forall \  \mathcal{D}(t') \in \mathcal{D} \ \text{and for} \ \psi \in \mathcal{R} \, ,
\end{equation}
where $\mathbb{I}(\cdot)$ is an indicator function. In allocating spells to each $t'$, we shall use the `cohort-end'-definition of reporting time from \citet{botha2025recurrentEvents}, such that each $\mathcal{D}(t')$-set contains all spells $(i,j)$ that commonly stop at a given $t'$-value.

\begin{figure}[ht!]
    \centering
    \includegraphics[width=0.8\linewidth, height=0.47\textheight]{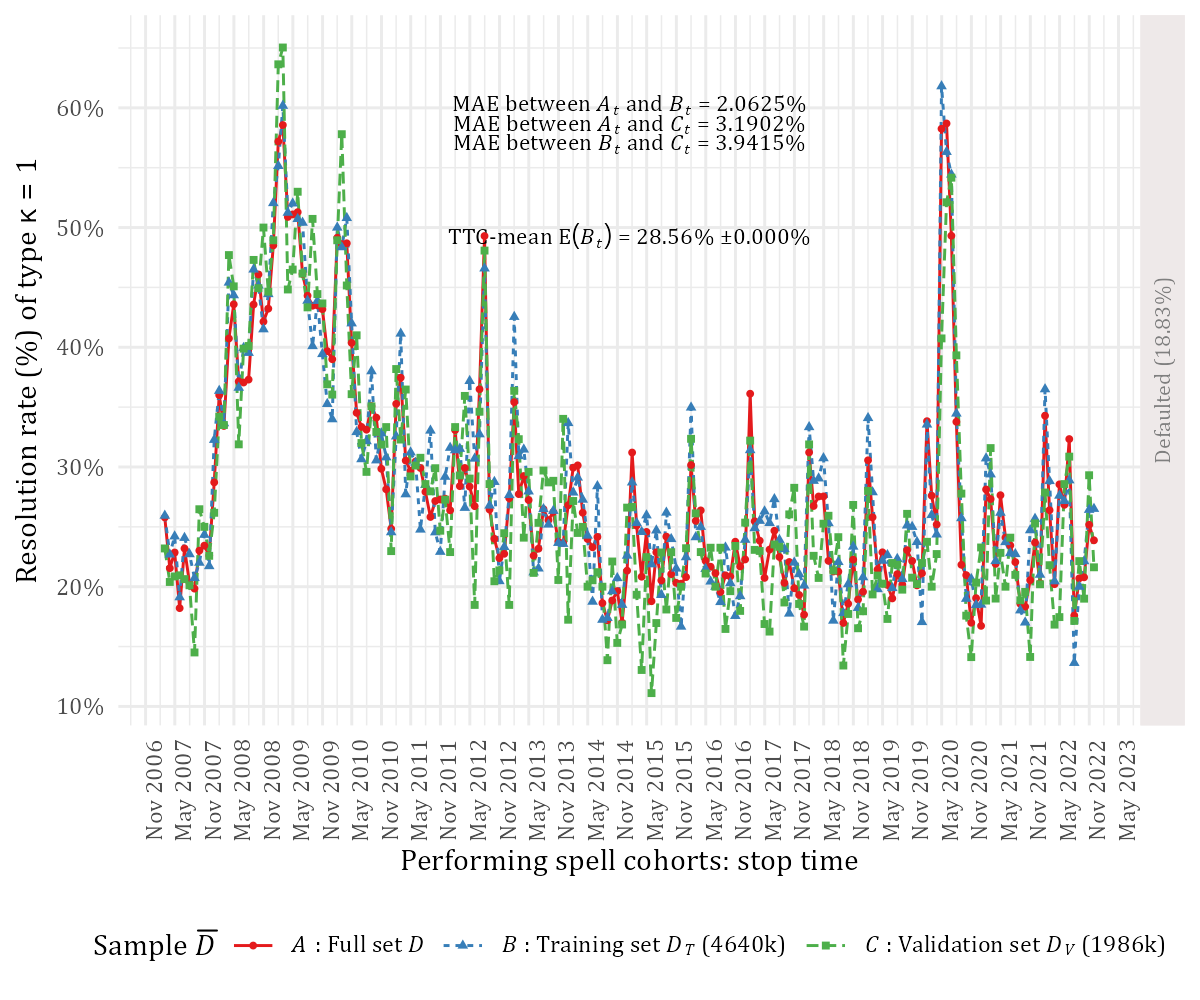}
    \caption{Comparing the resolution rates of type $\psi=1$ (Default) over time across the various datasets, as sampled from residential mortgage data. The MAE-based AD-measure from \autoref{eq:ResolRate_MAE} summarises the discrepancies over time for each dataset-pair.}
    \label{fig:ResolRate_PWP}
\end{figure}

The sets $\left\{ \mathcal{D}_T, \mathcal{D}_V \right\}$ can now be checked for time-dependent sampling bias using \autoref{eq:ResolRate}. As in \citet{botha2025recurrentEvents}, we duly calculate the resolution rates $r_\psi(t',\mathcal{D}_T)$ and $r_\psi(t',\mathcal{D}_V)$, whereafter they are screened for large discrepancies over $t'$. 
Such discrepancies can be detected using the absolute difference, $|r_\psi(t',\mathcal{D}_T)$ - $r_\psi(t',\mathcal{D}_V)|$, which should be as close to zero as possible, thereby affirming both datasets to be representative of each other. The mean hereof, i.e., the \textit{mean absolute error} (MAE), is used in defining the \textit{average discrepancy} (AD) as a diagnostic measure within our context. We express this AD-measure over $t'$ between any two non-overlapping sets $\mathcal{D}_1$ and $\mathcal{D}_2$ as
\begin{equation} \label{eq:ResolRate_MAE}
     \text{AD: } \quad \bar{r}_\psi\left(\mathcal{D}_1, \mathcal{D}_2 \right) = \frac{1}{n} \sum_{t'}{\big\vert r_\psi(t',\mathcal{D}_1) - r_\psi(t',\mathcal{D}_2) \big\vert} \quad \forall \ t' \ \text{and for} \ \psi \in \mathcal{R}\, .
\end{equation}
This AD-measure can be computed for all combinations of the datasets $\left\{ \mathcal{D}, \mathcal{D}_T, \mathcal{D}_V \right\}$, thereby resulting in the collection $\left\{ \Bar{r}_\psi(\mathcal{D},\mathcal{D}_T), \Bar{r}_\psi(\mathcal{D},\mathcal{D}_V), \Bar{r}_\psi(\mathcal{D}_T,\mathcal{D}_V) \right\}$. In \autoref{fig:ResolRate_PWP}, we provide an example of comparing the default resolution rate ($\psi=1$) across the different datasets. Clearly, the rates track both the 2008 financial crisis and the Covid-2019 pandemic. More importantly, the AD-measure complements a visual analysis in that all rates are reasonably close to one another, with an AD-value of about 2\% between $\mathcal{D}$ and $\mathcal{D}_T$. These diagnostic results suggest that the resampled sets exhibit only a low degree of sampling bias over time, which bodes well for the eventual model's ability to generalise beyond training data.

%% file: 3.3-Fundamentals.tex
\subsection{Constructing the empirical term-structure of default risk within discrete-time}
\label{sec:survival_fundamentals}

Though time is continuous in its very nature, it is often measured in discrete intervals. Likewise for the credit domain, loan performance is observed only at the end of each month, even though default and competing events can technically occur anytime during the month. For this reason, the discretisation of time within the credit domain poses no loss of information.
From \citet{singer1993time}, \citet[pp.~15-17,~\S 4.1]{jenkins2005survival}, \citet[\S7]{allison2010survival}, \citet[pp.~15-16,~57-58,~81-82]{crowder2012credit}, \citet{kartsonaki2016survival}, and \citet{suresh2022survival}, it is said that continuous time can be discretised into a sequence of distinct contiguous intervals. This sequence of \textit{unique ordered failure times} (excluding censored cases) is defined as $\left(0,t_{(1)}\right], \left(t_{(1)},t_{(2)}\right], \dots, \left(t_{(k-1)},t_{(k)}\right], \dots \left(t_{(m-1)},t_{(m)}\right]$, where $t_{(1)} < t_{(2)},< \dots < t_{(k)} < \dots < t_{(m)}$ up to some maximum value $m$.
The $k^\mathrm{th}$ period starts immediately after the point $t_{(k-1)}$, whereafter the interval ends precisely at $t_{(k)}$, both of which can coincide with the start and end of each month in credit data. Within each of these intervals $\left({t_{(k-1)}}, {t_{(k)}}\right]$, one may record the occurrence or non-occurrence of events in aggregate. In particular, let $c_k$ denote the number of spells that are right-censored during said interval, and let $f_k$ represent the number of spells that have failed/defaulted at $t_{(k)}$. Both of these quantities are observed at month-end in credit data. Then, $n_k$ is the number of spells that are \textit{at risk} of ending immediately prior to $t_{(k)}$, where such spells have a spell age of at least $T_{ij}>t_{(k)}$, and reside within the risk set. One can express $n_k$ as the summation over all failure times $f_q$ and censoring times $c_q$, where $q\geq k$ indexes those remaining times from and beyond $t_{(k)}$, i.e.,
\begin{equation}\label{eq:risk_set}
    n_k = \left( f_q + c_q \right) + \left( f_{q+1} + c_{q+1} \right) + \dots + \left( f_m + c_m \right) =     
    \sum_{q=k }^{m} (f_q + c_q) \, , 
\end{equation}
assuming that $f_0=0$ and that $n_0$ is the initial population count.
Furthermore, the aforementioned event history indicator $e_{ijt}$ from \autoref{sec:survival_concepts} can be expressed using these intervals with an indicator function; i.e., $e_{ijk}=\mathbb{I}\left(t_{(k-1)} <T_{ij} \leq t_{(k)}\right)$, which equals 1 if subject-spell $(i,j)$ defaulted during $\left(t_{(k-1)},t_{(k)} \right]$, and 0 otherwise.

The lifetime $T_{ij}$ of each spell $(i,j)$ is now considered to be a realisation from a non-negative discrete random variable $T$. From \citet[pp.~17-20]{jenkins2005survival}, \citet[pp.~15-16]{crowder2012credit}, and \citet{suresh2022survival}, let $F\left(t_{(k)}\right)=\mathbb{P}\left(T \leq t_{(k)}\right)$ denote the cumulative lifetime distribution, which evaluates the probability of experiencing default within a specified time horizon $(t_{(0)}, t_{(k)}]$; i.e., the \textit{cumulative default probability distribution}. Its complement $S\left(t_{(k)}\right) = 1-F\left( t_{(k)} \right) = \mathbb{P}\left(T > t_{(k)}\right)$ is the classical survivor function, and it has an associated probability mass function $f\left( t_{(k)} \right)=\mathbb{P}\left( T=t_{(k)} \right)$, which is also the \textit{marginal default probability} of $T$ assuming a particular event time; see \citet[p.~190]{baesens2016credit} and \citet{bank2021review}. In discrete-time, the marginal PD $f\left(t_{(k)}\right)$ is related to $S\left(t_{(k)}\right)$ as
\begin{align}\label{eq:survivor_discTime} 
    S\left(t_{(k)}\right) &= \sum_{s: \, t_{(s)} \, > \, t_{(k)}}f\left(t_{(s)}\right) \, , \quad \text{with} \\ 
    f\left(t_{(k)} \right) &= \mathbb{P}\left( T=t_{(k)} \right) = \mathbb{P}\left( t_{(k-1)} < T \leq t_{(k)} \right) = S\left(t_{(k-1)} \right) - S\left(t_{(k)} \right) \, . \label{eq:PDF_discTime} 
\end{align}
It is commonly assumed that zero-length lifetimes are not possible such that $S\left(t_{(0)}\right)=1$, while $f(t)=0$ whenever $t$ does not equal any of the ordered failure times $t_{(k)}$.

Another useful quantity in survival analysis is the \textit{discrete hazard} $h\left( t_{(k)} \right)$, which is the proportion of the risk set prior to $t_{(k)}$ that has experienced default during the contiguous interval $\left({t_{(k-1)}}, {t_{(k)}} \right]$, i.e., the \textit{conditional default probability}. More formally, and in following \citet[pp.~17-20]{jenkins2005survival} and \citet[pp.~15-16]{crowder2012credit}, this $h\left( t_{(k)} \right)$-quantity becomes the conditional probability of exiting the spell during the $k^\mathrm{th}$ interval, having survived hitherto, and is expressed as
\begin{equation} \label{eq:hazardfunction_discTime}
    h\left(t_{(k)}\right) = \mathbb{P}\left(  t_{(k-1)} < T \leq t_{(k)} \, | \, T > t_{(k-1)} \right) = 1 - \frac{S\left(t_{(k)} \right)}{S\left(t_{(k-1)} \right)} \, .
\end{equation}
Note that $0 \leq h\left( t_{(k)} \right) \leq 1$ for all $t$ in discrete time, but $h(0)=0$ and $h\left(t_{(m)}\right)=1$ only at the last event time $t_{(m)}$. Perhaps more importantly, the marginal PD $f\left( t_{(k)} \right)$ is related to the conditional PD $h\left( t_{(k)} \right)$ as
\begin{equation} \label{eq:discHaz_eventProb}
    f\left( t_{(k)} \right) = S\left( t_{(k-1)} \right)\cdot h\left( t_{(k)} \right) \quad \implies \quad h\left( t_{(k)} \right)=\frac{f\left( t_{(k)} \right)}{S\left( t_{(k-1)} \right)}\,.
\end{equation}
In our context, it is exactly the collection $f(t),t=t_{(0)},\dots,t_{(m)}$ that constitutes the \textit{term structure} of default risk.

In estimating these survival quantities, note that the survivor function $S\left( t_{(k)} \right)$ can also be expressed as a multiplicative chained sequence of discrete hazards, i.e., 
\begin{equation} 
    S\left( t_{(k)} \right) = \prod_{s \, : \, t_{(s)} \, \leq \, t_{(k)}}(1-h\left( t_{(s)} \right) \, . \label{eq:survfunct_discrete_chained_general}
\end{equation}
Given the aforementioned failure counts $f_k$ and at-risk counts $n_k$ from \autoref{eq:risk_set}, \citet[pp.~15,~55,~77,~81]{crowder2012credit} and \citet{kartsonaki2016survival} showed that a particular discrete hazard may be estimated by setting $\hat{h}\left( t_{(k)} \right)=h_k=f_k/n_k$. This result naturally leads to the well-known \textit{Kaplan-Meier} (KM) estimator from \citet{kaplan1958credit}, defined as
\begin{equation} \label{eq:KaplanMeier}
   \hat{S}\left( t_{(k)} \right)=\prod_{s \, : \, t_{(s)} \, \leq \, t_{(k)}}{\left( 1-\frac{f_s}{n_s} \right)} = \prod_{s \, : \, t_{(s)} \, \leq \, t_{(k)}}{\left( 1-h_s \right)} \, .
\end{equation}

\begin{figure}[!ht]
    \centering
    \includegraphics[width=0.85\linewidth,height=0.6\textheight]{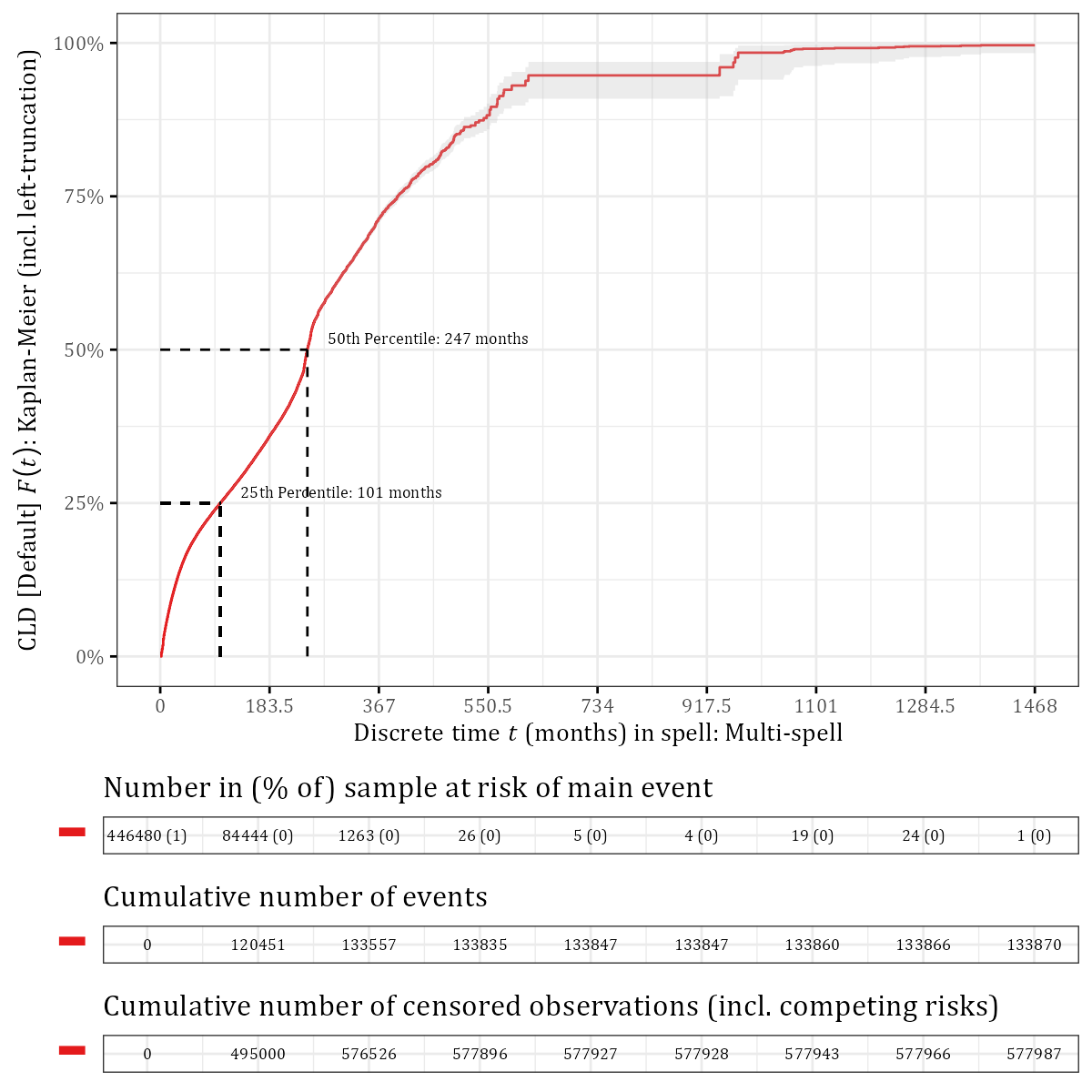}
    \caption{Illustrating the cumulative lifetime distribution (CLD) of the time to default, having used residential mortgage data with the KM-estimator from \autoref{eq:KaplanMeier}. Greenwood's formula is used in calculating 95\% confidence intervals, as implemented in the \texttt{ggsurvplot()}-function within the \texttt{survminer} R-package.}
    \label{fig:KaplanMeier_CLD}
\end{figure}

In \autoref{fig:KaplanMeier_CLD}, we illustrate the KM-estimator from \autoref{eq:KaplanMeier} using the same residential mortgage data. However, it is of greater interest to graph the cumulative default distribution $F(t)=1-S(t)$ for $t=t_{(0)},\dots,t_{(m)}$, given our focus on risk parameters. In \autoref{fig:KaplanMeier_CLD}, the widening confidence interval, itself calculated using Greenwood's formula, suggests diminishing sample sizes at later default times. A visual analysis can aid in finding an inflexion point beyond which the sample size would dwindle too much, thereby possibly compromising the modelling process later. From the example, it seems that specifying $t\leq 300$ would safely discard the more extreme spell ages, particularly given the profile of the underlying 20-year mortgage portfolio.
Moreover, the accompanying risk table in \autoref{fig:KaplanMeier_CLD} confirms the visual analysis since the underlying dataset is already exhausted at this particular inflection point, i.e., approximately 0\% remain at risk. Regarding the construction of \autoref{fig:KaplanMeier_CLD}, refer to script 4b in the R-codebase of \citet{botha2025termStructureSourcecode}.

\begin{figure}[!ht]
    \centering
    \includegraphics[width=0.7\linewidth,height=0.43\textheight]{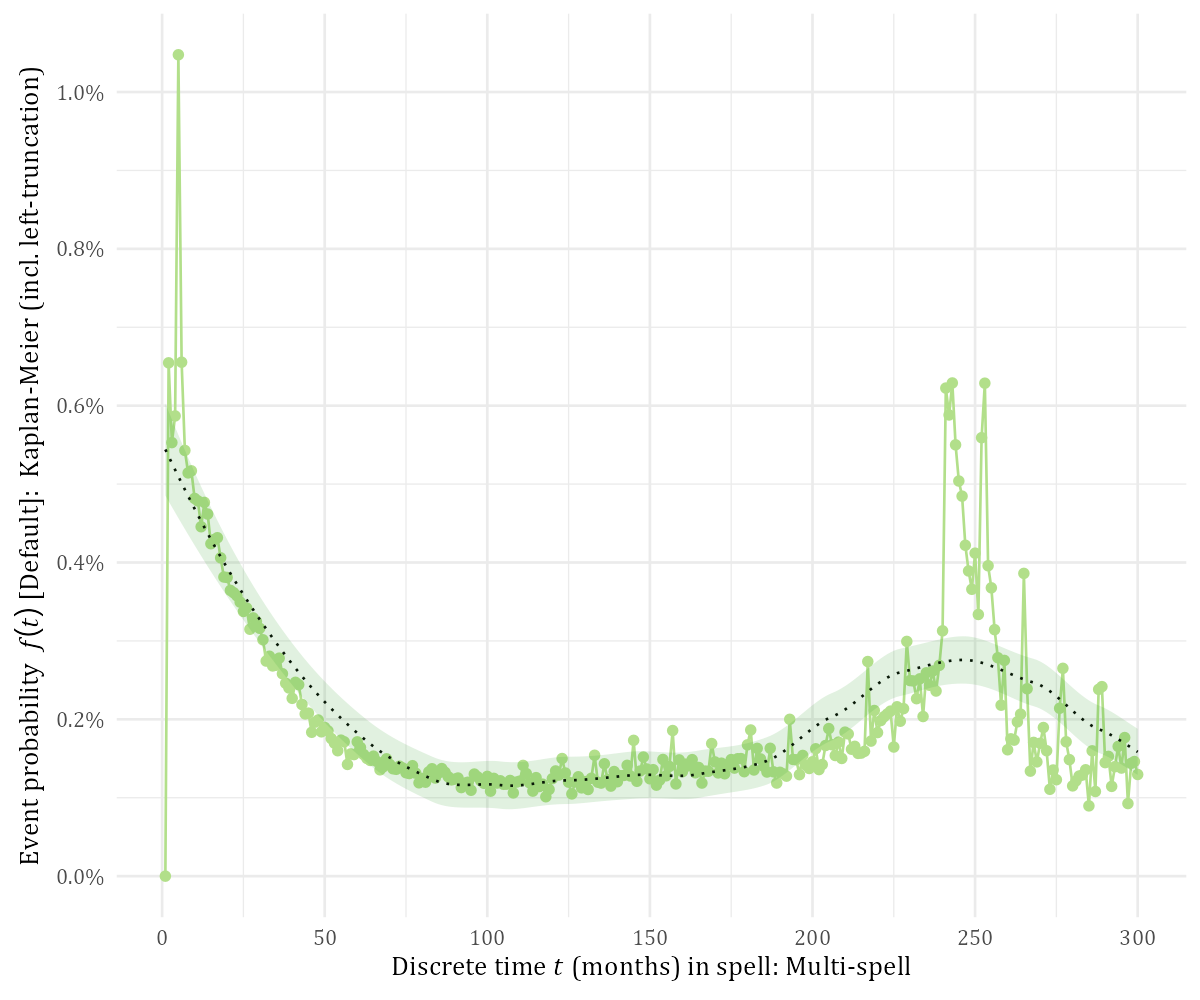}
    \caption{The empirical term-structure of default risk, or marginal PDs, as constituted by the discrete event probabilities $f(t),t=t_{(1)},\dots,t_{(m)}$ over spell time. Its estimation relies upon the KM-estimator from \autoref{eq:KaplanMeier} using residential mortgage data. A LOESS-smoother with a 95\% confidence interval is overlaid merely to summarise the visual trend.}
    \label{fig:KaplanMeier_EventProb}
\end{figure}

Following the estimation of $S(t),t=t_{(0)},\dots,t_{(m)}$ using \autoref{eq:KaplanMeier}, one can derive perhaps the most important quantity in survival analysis, at least within the context of credit risk modelling. That is, we shall estimate the empirical event probabilities (or marginal default probabilities) using \autoref{eq:PDF_discTime}, which constitutes the term-structure of default risk, i.e., the collection $\left\{ f\left(t_{(k)} \right) \right\}_{k=1}^{m}$. Using the same data, we illustrate in \autoref{fig:KaplanMeier_EventProb} these event probabilities over time and note that it has the characteristic U-shape: high at first, then down for while, then up again. The hazard rate also exhibits this U-shape and, as \citet[p.~14]{crowder2012credit} explained, it attests of a system that is prone to wear and tear over time. A U-shaped hazard rate typically manifests in survival applications within the manufacturing domain as well, e.g., the lifetimes of new light bulbs. Such units typically have a high risk of early failure ("wear in"), as well as a high risk of late failure ("wear out"). Likewise, early-dated loans have greater default risk than usual, likely due to new home owners still contending with the ordeal of repaying a mortgage. The greater default risk of later-dated loans are ascribed to a deluge of older left-truncated subject-spells that defaulted soon after the 2008 financial crisis. Another explanation is that some borrowers opt for strategic default whilst in the process of selling a property, which is somewhat more prevalent towards the end of a mortgage term than at its start.
Nonetheless, this empirical term-structure can be used as an outcome variable of sorts against which model-driven predictions can be compared in aggregate. In short, this KM-derived construct in \autoref{fig:KaplanMeier_EventProb} serves as the "actual"-part of a typical "actual vs expected" comparative setup in measuring the calibration success of any particular model of the term-structure, as we shall demonstrate later.

%% file: 3.4-DiscTimeCoxModel.tex
\subsection{A discrete-time hazard model using the Prentice-Williams-Peterson (PWP) technique}
\label{sec:survival_discTimeCoxModel}

One can risk-sensitise the various survival quantities from \autoref{sec:survival_fundamentals} by conditioning them on a set of inputs (or covariates) $\boldsymbol{x}_{ij}$ that are specific to the subject-spell $(i,j)$. Note that the fundamental relationships amongst the various survival quantities, or $f(t|\,\boldsymbol{x}_{ij}),h(t|\,\boldsymbol{x}_{ij}),S(t|\,\boldsymbol{x}_{ij})$, remain unchanged. In preparation for modelling, let us briefly consider the underlying likelihood of observing the data in discrete-time. Firstly, let $m_{ij}$ denote the last/terminal interval during which we still have any information about $(i,j)$, such that the spell age is $T_{ij}\in\left( t_{(m_{ij}-1)} ,t_{(m_{ij})} \right]$. 
Because of censoring, \citet{singer1993time} and \citet{suresh2022survival} stated that the likelihood then consists of two types of contributions: 1) for uncensored spells, the probability of defaulting during the interval $m_{ij}$; and 2) for censored spells, the probability of defaulting after $m_{ij}$, i.e., the survival probability.
As such, let the spell-level failure indicator $\delta_{ij}\in\{0,1\}$ equal 1 if $(i,j)$ defaulted during $\left( t_{(m_{ij}-1)} ,t_{(m_{ij})} \right]$, and zero otherwise. Subject-spells with $\delta_{ij}=1$ therefore contribute the failure probability to the likelihood, and do not contribute any further information beyond $t_{(m_{ij})}$. We may then express the likelihood function $L$ as
\begin{align} \label{eq:survival_likelihood}
    L &= \prod_{(i,j)} {\left[ \mathbb{P}\left(T_{ij}=t_{m_{ij}} \right) \right]^{\delta_{ij}} \left[ \mathbb{P}\left( T_{ij} > t_{m_{ij}} \right)  \right]^{1-\delta_{ij}}} \nonumber \\
    &= \prod_{(i,j)} \left[ h\left(t_{(m_{ij})} | \, \boldsymbol{x}_{ij} \right) S\left(t_{(m_{ij}-1)} | \, \boldsymbol{x}_{ij} \right)\right]^{\delta_{ij}} \left[ S\left(t_{(m_{ij})} | \, \boldsymbol{x}_{ij} \right)  \right]^{1-\delta_{ij}} \nonumber \\
    &= \prod_{(i,j)} \left[ h\left(t_{(m_{ij})} | \, \boldsymbol{x}_{ij}\right) \left(\prod_{q=1}^{m_{ij}-1}{ 1- h\left(t_{(q)} | \, \boldsymbol{x}_{ij}\right) } \right)  \right]^{\delta_{ij}} \left[ \prod_{q=1}^{m_{ij}-1}{ 1- h\left(t_{(q)} | \, \boldsymbol{x}_{ij}\right) } \right]^{1-\delta_{ij}}  \, .
\end{align}

As derived by \citet{singer1993time} and \citet{suresh2022survival}, it turns out that $L$ can be rewritten as
\begin{equation} \label{eq:survival_likelihood_final}
    L=\prod_{(i,j)}\prod_{q=1}^{m_{ij}}{ h\left(t_{(q)}| \, \boldsymbol{x}_{ij}\right)^{e_{ijq}}\left( 1 - h\left(t_{(q)} | \, \boldsymbol{x}_{ij}\right) \right)^{1-e_{ijq}}  } \, ,
\end{equation}
where $e_{ijk}$ is the event history indicator from \autoref{sec:survival_concepts}, defined as $e_{ijk}=\mathbb{I}\left(t_{(k-1)} <T_{ij} \leq t_{(k)}\right)$ for the $k^\mathrm{th}$ interval of subject-spell $(i,j)$.
This likelihood function is equivalent to that of a sequence of independent Bernoulli trials with parameters $h\left(t_{(q)}| \, \boldsymbol{x}_{ij}\right)$, i.e., the binomial model. In turn, the observed event indicators $e_{ijt},t=1,\dots,m_{ij}$ can be treated as a collection of independent dichotomous random variables. \citet{cox1972regression} originally proposed that these variables can have a logistic dependence on both the input variables and the spell periods (or intervals). 
However, \citet{suresh2022survival} noted that any algorithm that can optimise a binomial log-likelihood can be used in fitting a DtH-model. One can therefore use any statistical learning method that outputs the probability of a binary event, ranging from classical regression methods (e.g., GLMs) to machine learning approaches.

In the interest of simplicity, we shall opt for a GLM-framework and choose a logit link function in developing this tutorial, i.e., logistic regression. As such, and in following \citet{singer1993time}, our DtH-model is specified as
\begin{equation} \label{eq:discTimeHazard_logit}
    h\left(t| \, \boldsymbol{E}_{ij}, \boldsymbol{x}_{ij}, \boldsymbol{x}_{ij}(t), \boldsymbol{x}(t) \right) = \frac{1}{ 1 + \exp{\left( -\left[ \boldsymbol{\alpha}^\mathrm{T} \boldsymbol{E}_{ij} +  \boldsymbol{\beta}^\mathrm{T}\boldsymbol{x}_{ij} + \boldsymbol{\gamma}^\mathrm{T} \boldsymbol{x}_{ij}(t) + \boldsymbol{\delta}^\mathrm{T} \boldsymbol{x}(t) \right] \right) } } \, .
\end{equation}
In \autoref{eq:discTimeHazard_logit}, the vector $\boldsymbol{E}_{ij}=\left\{E_{ij1},\dots,E_{ijm} \right\}$ contains indicator variables that flag a specific period $t\in\left[1,\dots,m\right]$ during the discretely-valued history of a subject-spell $(i,j)$, up to the observed maximum $m$. These period indicators are accompanied by the estimable coefficients $\boldsymbol{\alpha}=\left\{\alpha_1,\dots,\alpha_m\right\}$; which together with the period indicators, form the baseline hazard. No single stand-alone intercept therefore exists. Furthermore, $\boldsymbol{\beta}=\left\{\beta_1,\dots,\beta_p \right\}$ are estimable coefficients for the $p$ spell-level time-fixed input variables, denoted by  $\boldsymbol{x}_{ij}=\left\{ x_{ij1},\dots,x_{ijp}\right\}$, where these inputs are measured at the start of each subject-spell. Similarly, the regression coefficients $\boldsymbol{\gamma}=\left\{ \gamma_1, \dots, \gamma_{p'} \right\}$ and $\boldsymbol{\delta}=\left\{ \delta_1, \dots, \delta_{p^*} \right\}$ respectively accompany the $p'$ time-dependent variables $\boldsymbol{x}_{ij}(t)=\left\{x_{ij1}(t),\dots,x_{ijp'}(t) \right\}$ of each $(i,j)$, and the $p^*$ portfolio-level time-dependent variables $\boldsymbol{x}(t)=\left\{ x_1(t), \dots, x_{p^*}(t) \right\}$, e.g., macroeconomic variables.

When compared to the CPH-model from \citet{cox1972regression}, the DtH-model avails at least four benefits, as discussed by \citet{suresh2022survival} and \citet[\S7]{allison2010survival}. 
Firstly, the proportional hazards assumption is no longer required, which is itself unlikely to hold in practice. 
Secondly, the DtH-model allows for a more intuitive interpretation of `hazard' as the event probability during a certain time interval, which is conditioned on first surviving to said interval. 
Thirdly, the DtH-model can easily handle tied event times without needing adjustments, which would have been the case with CPH-models. In this regard, \citet[\S 7]{allison2010survival} explained that the exact methods for handling such ties within a CPH-model are computationally demanding, whereas the available approximation methods (e.g., Breslow's estimator) perform poorly in heavily tied data.
Lastly, a DtH-model avails explicit coefficient estimates for the baseline hazard function at each interval. These estimates are not usually available within a CPH-model, which can complicate prediction tasks when using such a CPH-model.

In fitting the DtH-model from \autoref{eq:discTimeHazard_logit} onto the training dataset $\mathcal{D}_T$ from \autoref{tab:dataStructure_perfSpells_PWP}, itself denoted by the object \texttt{datTrain}, we shall use the well-known \texttt{glm()} function from the \texttt{stats}-package in the R-programming language. The spell period $t_{ij}$ tracks the time spent in the performing spell $(i,j)$ at each row of its duration, where $t_{ij}$ is denoted by the field \texttt{SpellPeriod}. In essence, we regress the event history indicators $e_{ijt}$ over $t_{ij}=\tau_e,\dots,\tau_s$, as represented by the field \texttt{Event}, onto the entire input space. This set of input variables includes application, behavioural, and macroeconomic variables; as well as the spell period $t_{ij}$, itself discretised using the \texttt{factor()}-statement. We program\footnote{In the SAS-programming language, \citet{allison2010survival} showed that the standard errors of coefficients produced by the PROC PHREG procedure are virtually identical to those produced by PROC LOGISTIC.} this operation as follows.
\begin{lstlisting}
modLR <- glm(Event ~ -1 + factor(SpellPeriod) + Inputs, data=datTrain, family="binomial")
\end{lstlisting}

There are certainly other more parsimonious ways of embedding the baseline hazard over time into the model, other than via dummy-encoding. These ways include incorporating $t_{ij}$ either directly with a single coefficient $\alpha_1$, via a function such as $\log{t_{ij}}$, by using a regression spline function as in \citet{djeundje2019dynamic}, or via a binning scheme that gives a reduced set of time indicator variables $E'_1,\dots,E'_{m'}$. As shown below using the \texttt{dplyr} and \texttt{data.table} R-packages, we have opted for a binning scheme. Adopting such a scheme achieved a good balance between capturing the non-linear shape of the baseline hazard, and model parsimony. Naturally, other binning schemes could be more appropriate for other portfolios.
\begin{lstlisting}
timeBinning <- function(x) {
  case_when(
    0 < x & x <= 3 ~ "01.[1,3]", 3 < x & x <= 6 ~ "02.(3,6]",
    6 < x & x <= 9 ~ "03.(6,9]", 9 < x & x <= 12 ~ "04.(9,12]",
    12 < x & x <= 18 ~ "05.(12,18]", 18 < x & x <= 24 ~ "06.(18,24]",
    24 < x & x <= 30 ~ "07.(24,30]", 30 < x & x <= 36 ~ "08.(30,36]",
    36 < x & x <= 48 ~ "09.(36,48]", 48 < x & x <= 60 ~ "10.(48,60]",
    60 < x & x <= 72 ~ "11.(60,72]", 72 < x & x <= 84 ~ "12.(72,84]",
    84 < x & x <= 96 ~ "13.(84,96]", 84 < x & x <= 96 ~ "14.(84,96]",
    96 < x & x <= 108 ~ "15.(96,108]", 108 < x & x <= 120 ~ "16.(108,120]",
    120 < x & x <= 144 ~ "17.(120,144]", 144 < x & x <= 168 ~ "18.(144,168]",
    168 < x ~ "19.168+",TRUE ~ "20.168+"
  )
}
datTrain[, Time_Bn := timeBinning(SpellPeriod)]
\end{lstlisting}

In dealing with recurrent default events, we adopt a specific approach called the \textit{Prentice-Williams-Peterson} (PWP) technique. This PWP-technique extends the common CPH-model, as discussed and illustrated by \citet{amorim2015modelling}, \citet{ozga2018systematic}, and \citet{botha2025recurrentEvents}. The Cox-regression model is itself discussed at length by \citet[\S 3.1]{therneau2000modeling}, \citet[\S 4.2]{crowder2012credit}, and \citet{schober2018survival}, and so we shall not further contend with it here.
More importantly, the PWP-technique analyses the ordering of spells and stratifies the data based on the spell number $j$. This process involves fitting a spell-specific baseline hazard function, thereby accounting for changes in the baseline risk between two successive spells. 
In our context, the PWP-technique implies fitting an interaction term amongst the series of time indicators $E'_1,\dots,E'_{m'}$, and the spell number $j=1,\dots,J$, where $J$ is the maximum observed (binned) spell number. We reprogram the previous operation as follows, where the field \texttt{SpellNum\_Bn} represents a binned version of $j$, limited to four bins ("1", "2", "3", "4+").
\begin{lstlisting}
modLR <- glm(Event ~ -1 + Time_Bn*SpellNum_Bn + Inputs, data=datTrain, family="binomial")
\end{lstlisting}

Since the default-event is a rare outcome, our sampled training dataset is said to be imbalanced. Depending on its prevalence, class imbalance may compromise the performance of a model on unseen data. One of the strategies by which this imbalance can be addressed is by modifying the cost associated with misclassification. As discussed by \citet{he2009learning}, this \textit{cost-sensitive learning} approach can incorporate such costs either explicitly (via a cost matrix), or implicitly by weighing each observation differently during model training.
In the interest of expediency, we have opted for fitting our DtH-model using a \textit{weighted logistic regression} (WLR). This WLR-procedure remains within the GLM-framework with a binomial distribution and logit link, though its likelihood function now incorporates observation-specific weights, as reviewed by \citet{zeng2024comprehensive}. Following experimentation, we have obtained superior models when weighing/replicating defaulted observations 10 times more than non-defaults. This experimentation was based on minimising the MAE between the resulting 12-month expected default rate and its empirical counterpart, as confirmed in the upcoming \autoref{fig:DefaultRates} in \autoref{sec:survival_diagnostics}. More formally, those instances where $e_{ijt}=1$ will have a weight of $w_{ijt}=10$, as denoted by the field \texttt{Weight}; and $w_{ijt}=1$ elsewhere. We refit our DtH-model using these weights as follows.
\begin{lstlisting}
datTrain[, Weight := ifelse(Event==1,10,1)]
modLR <- glm(Event ~ -1 + Time_Bn*SpellNum_Bn + Inputs, data=datTrain, family="binomial", weights=Weight)
\end{lstlisting}

%% file: 3.5-Diagnostics.tex
\subsection{Various diagnostics in validating discrete-time hazard models}
\label{sec:survival_diagnostics}

The prevalence of right-censoring implies that evaluating the predictions and/or fit of a survival model is subject to the chosen time horizon over which predictions are made. Having chosen any such a time horizon, some in-sample observations will lack an outcome due to right-censoring, even though predictions are available. This means that most conventional model assessment tools would become inappropriate since they commonly assume that all observations are fully known, as discussed by \citet{Graf1999}, \citet{heagerty2000}, and \citet{suresh2022survival}.
For example, and since the time frame $t\geq 0$ can vary across which the event probability $f(t)$ is predicted, an ROC-based test of discriminatory power will also vary based on the degree of right-censoring at $t$. As formulated in \autoref{app:tDiagnostics}, one can however express the two elements of an ROC-graph as functions of $t$: the true positive rate $T^+(t)$ and the false positive rate $F^+(t)$.
As an illustration, we fit two competing DtH-models using residential mortgage data, which are respectively called the basic and advanced DtH-models. These models are differentiated by the breadth of their input spaces: six vs twelve input variables respectively; see \autoref{app:InputSpace} for details.

\begin{figure}[ht!]
\centering
\begin{subfigure}[b]{0.49\textwidth}
    \caption{Basic Dth-model}
    \centering\includegraphics[width=1\linewidth,height=0.3\textheight]{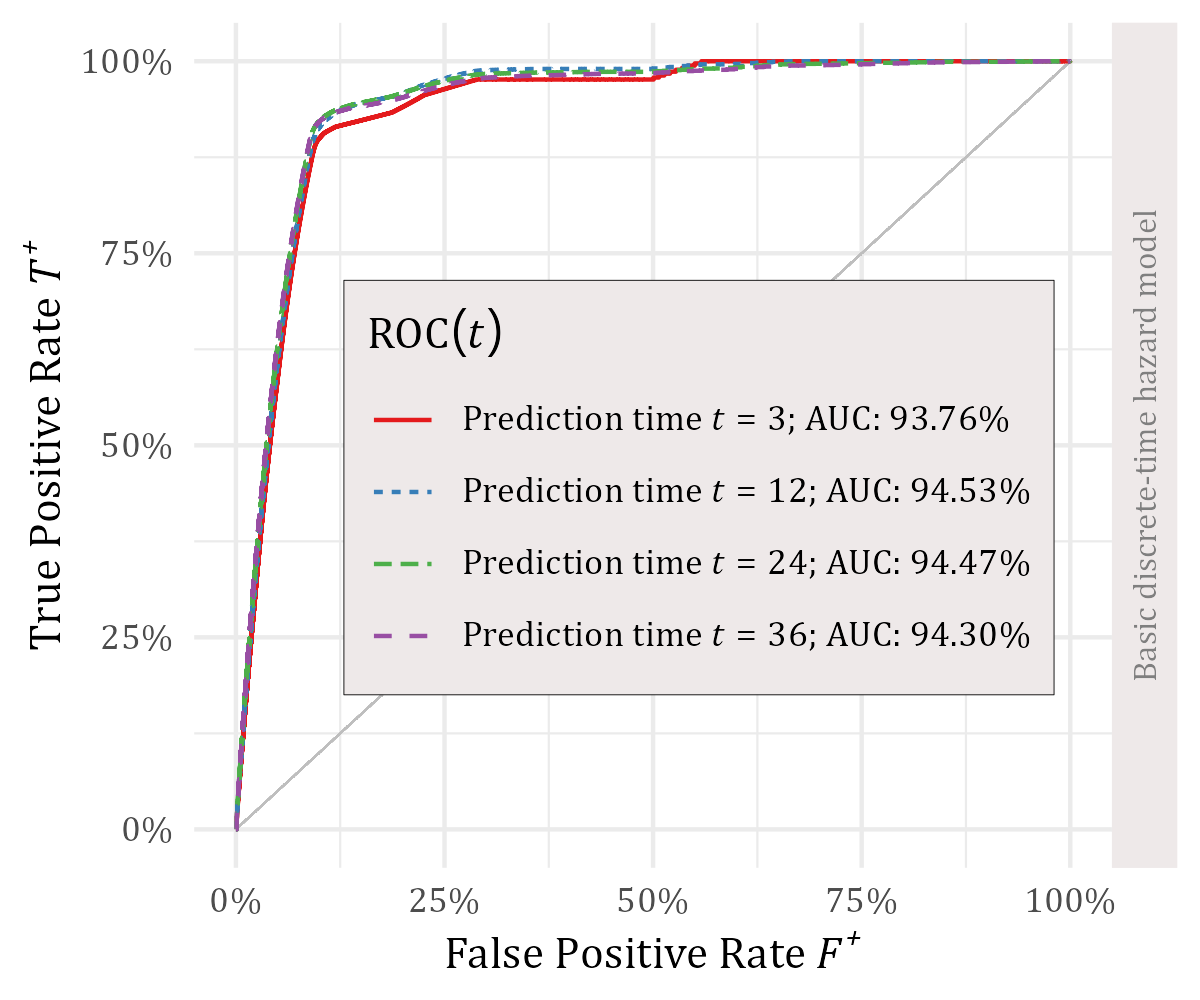}\label{fig:tROC_basic}
\end{subfigure} 
\begin{subfigure}[b]{0.49\textwidth}
    \caption{Advanced DtH-model}
    \centering\includegraphics[width=1\linewidth,height=0.3\textheight]{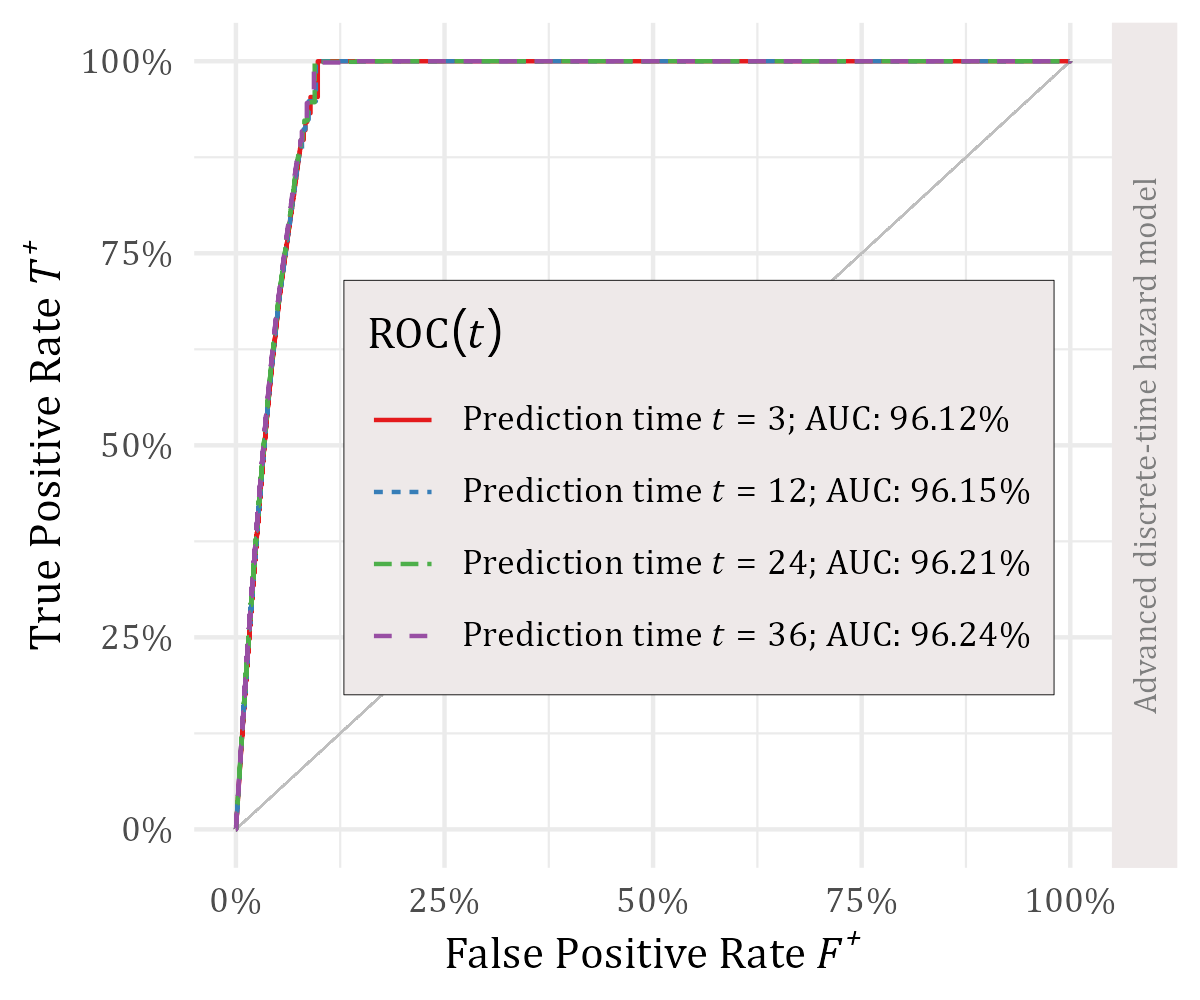}\label{fig:tROC_advanced}
\end{subfigure}%
\hfill%
\caption{Evaluating the discriminatory power of two competing discrete-time hazard (DtH) models by using the clustered tROC-extension to ROC-analysis at specific time points $t\in \{3,6,24,36\}$. The results per model are shown respectively in panels \textbf{(a)}-\textbf{(b)}, as compiled using the validation set $\mathcal{D}_V$.}\label{fig:tROC}
\end{figure}

As our first diagnostic in assessing the discriminatory power over time, we employ a \textit{time-dependent ROC-analysis} (tROC) across a small selection of time frames, denoted by $t\in\{3,12,24,36\}$ months spent in the performing spell; as shown in \autoref{fig:tROC}.
These tROC-results are then summarised into a set of \textit{time-dependent} AUC (tAUC) values, which are printed in \autoref{fig:tROC}; larger values indicate better model performance. For the construction of these graphs, we refer the reader to script 6a in the R-codebase of \citet{botha2025termStructureSourcecode}.
Overall, these tAUC-values are remarkably stable over $t$, except for the basic DtH-model wherein some larger differences are observed. Evidently, the discriminatory power of either model is excellent with tAUC-values commonly exceeding 90\%. We ascribe this result mainly to having embedded the full baseline hazard into the model using a binning scheme, which lends a great deal of predictive power. That said, tROC-analysis is known to be relatively (and demonstrably so) insensitive to evaluating improvements in the modelling technology, as shown by \citet{dirick2017time}. Nonetheless, the use of tROC-analyses remains a staple in the analytical toolbox of the analyst.

\begin{figure}[ht!]
\centering
\begin{subfigure}[b]{0.49\textwidth}
    \caption{Time-dependent Brier Scores}
    \centering\includegraphics[width=1\linewidth,height=0.28\textheight]{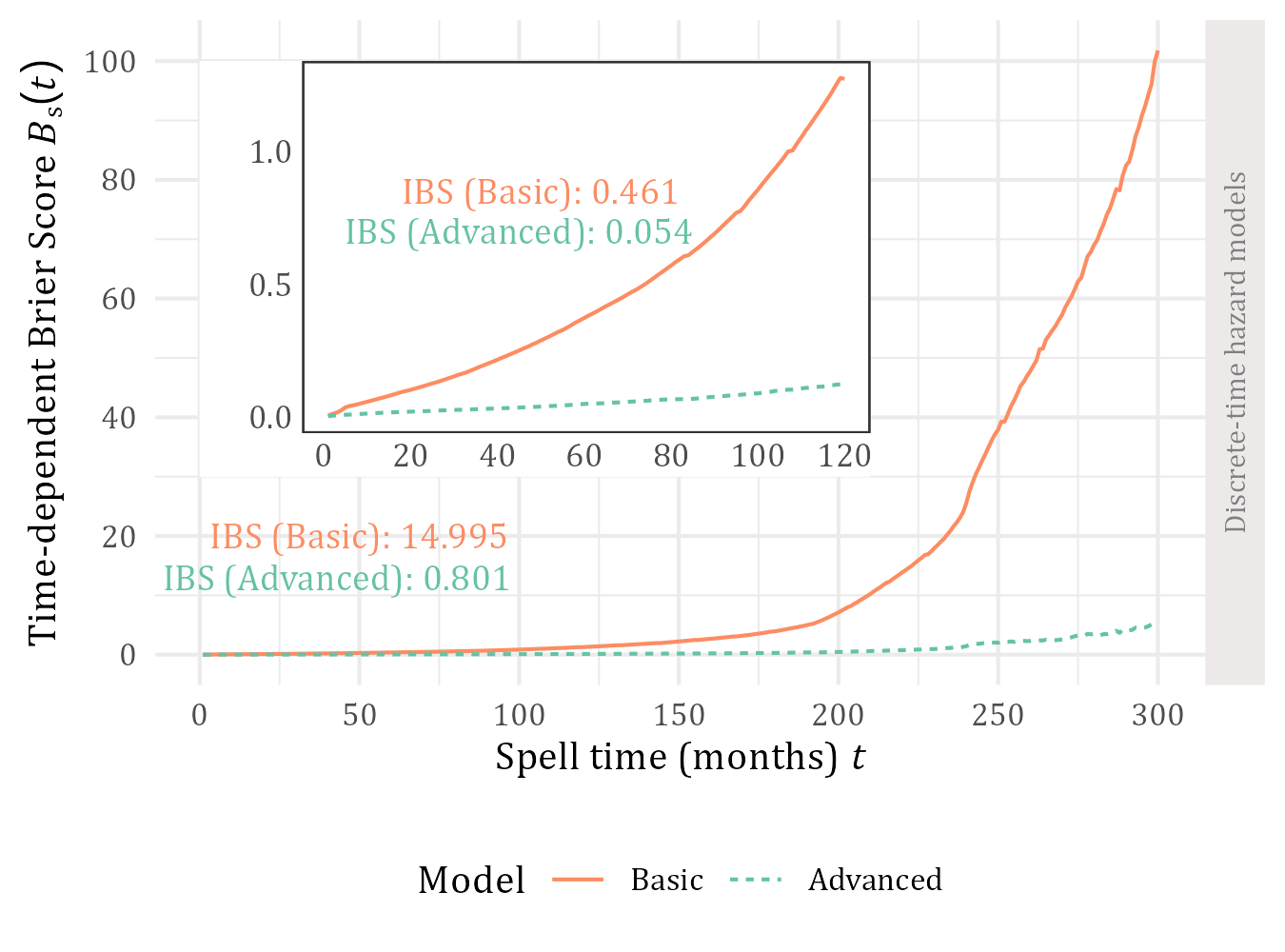}\label{fig:tBS}
\end{subfigure} 
\begin{subfigure}[b]{0.49\textwidth}
    \caption{Term-structures of default risk}
    \centering\includegraphics[width=1\linewidth,height=0.28\textheight]{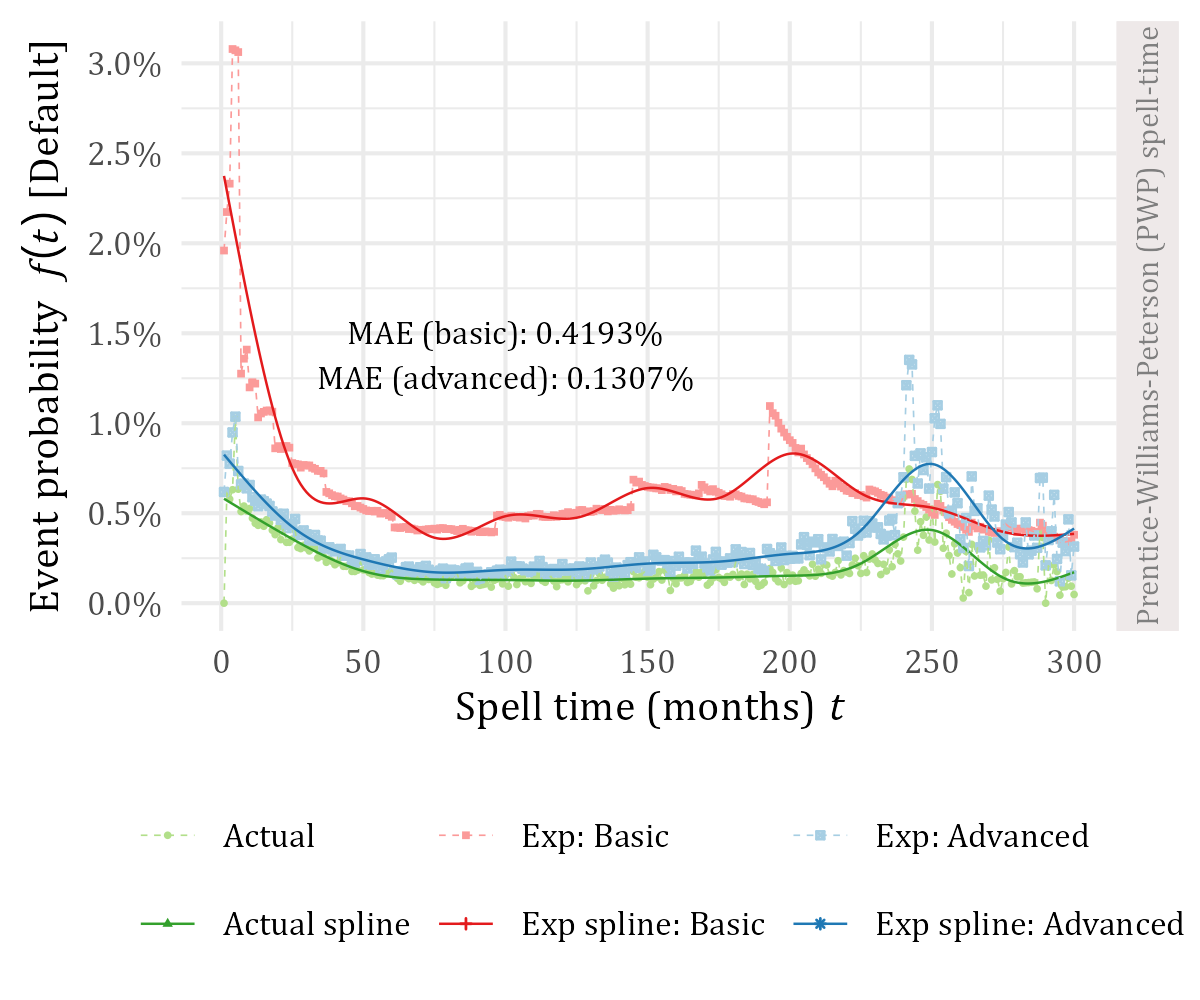}\label{fig:TermStructures}
\end{subfigure}%
\hfill%
\caption{Two quantities are shown over spell time $t$ on the x-asis. In \textbf{(a)}, the time-dependent Brier scores (tBS) are graphed for two competing DtH-models. In summarising the tBS-values, the integrated Brier score (IBS) is overlaid per model. The inset graph shows the same over a smaller time frame of 10 years. In \textbf{(b)}, the actual and expected ("exp") term-structures of default risk $f(t)$ are shown per model. A natural spline is overlaid (with 12 degrees of freedom) to summarise the visual trend. In summarising the average discrepancy between either model's term-structure and the empirical term-structure, the MAE is calculated and overlaid.}\label{fig:tbs_termStructure}
\end{figure}

As formulated in \autoref{app:tDiagnostics}, our second diagnostic is a \textit{time-dependent Brier score} (tBS). From \citet{Graf1999} and \citet{suresh2022survival}, this tBS-value measures both the discriminatory power and calibration (or prediction accuracy) of a survival model at a certain time frame $t\geq 0$. In particular, the tBS evaluates the average squared difference between predicted probabilities and observed outcomes at $t$ amidst right-censoring, where lower tBS-values indicate better model performance at $t$.
In \autoref{fig:tBS}, we graph the tBS-values across $t\leq 300$ months spent in the performing spell for both the basic and advanced DtH-models. Like the tAUC, these tBS-values can be summarised across $t$ into the \textit{integrated Brier score} (IBS), as discussed by \citet{Graf1999}, and printed in \autoref{fig:tBS}.
Moreover, the choice of the maximum period $t^*$ over which to calculate the tBS and IBS values should be meaningful, particularly amidst the increasing prevalence of right-censored cases as $t^*$ increases. Accordingly, we chose 10 years (120 months), as shown in the inset graph of \autoref{fig:tBS}.
Both DtH-models perform better at earlier time periods than later ones, though the degree of divergence between the two DtH-models is staggering as $t$ increases. Regarding IBS-values, we note the rudimentary rule of thumb from \citet{Graf1999}, who remarked that in the absence of information, one may plausibly assign a survival probability of $\hat{S}(t)=0.5$ to all subject-spells at a particular $t$. The corresponding empirical tBS would then be 0.25, which can be used as an upper cut-off when evaluating IBS-values from models; lower is better.
Evidently, the IBS-value for the advanced DtH-model (0.054) is orders of magnitude less than that of the basic DtH-model (0.461), especially over the 10-year time frame, which clearly positions the former as the superior model. The construction of this particular graph is programmed in script 6d of the accompanying R-codebase.

Following the empirical term-structure given previously in \autoref{fig:KaplanMeier_EventProb}, it is only natural to compare it to the model-driven term-structures over the unique ordered failure times $t=t_{(1)},\dots,t_{(m)}$, or spell time. Doing so would result in our third diagnostic, the \textit{term-structure comparison}, which measures the calibration quality between predictions and reality, as aggregated by spell time $t$. These expected term-structures of marginal PDs are constituted by the average event probability at each $t$, or the estimated $f\left(t\right)$ from \autoref{eq:PDF_discTime} and \autoref{eq:discHaz_eventProb}. We show these average event probabilities in \autoref{fig:TermStructures} respective to the basic and advanced DtH-models. The advanced DtH-model (blue) achieves a remarkably closer approximation to the empirical term-structure (green) than the basic DtH-model (red). In particular, the advanced DtH-model is capable of realising the expected U-shape in $f$ over $t$, whereas the basic DtH-model produces a term-structure with a much more jagged and disjoint shape. The quality of calibration is further measured using the MAE between the empirical and expected term-structures. Notably, the advanced DtH-model scored an MAE (0.1307\%) that is more than three times lower than that of the basic DtH-model (0.4193\%).
These results further support the positioning of the advanced DtH-model as the superior survival model. Lastly, we refer the interested reader to script 6c in the accompanying R-codebase regarding the estimation and graphing of these term-structures.

\begin{figure}[!ht]
    \centering
    \includegraphics[width=0.8\linewidth,height=0.5\textheight]{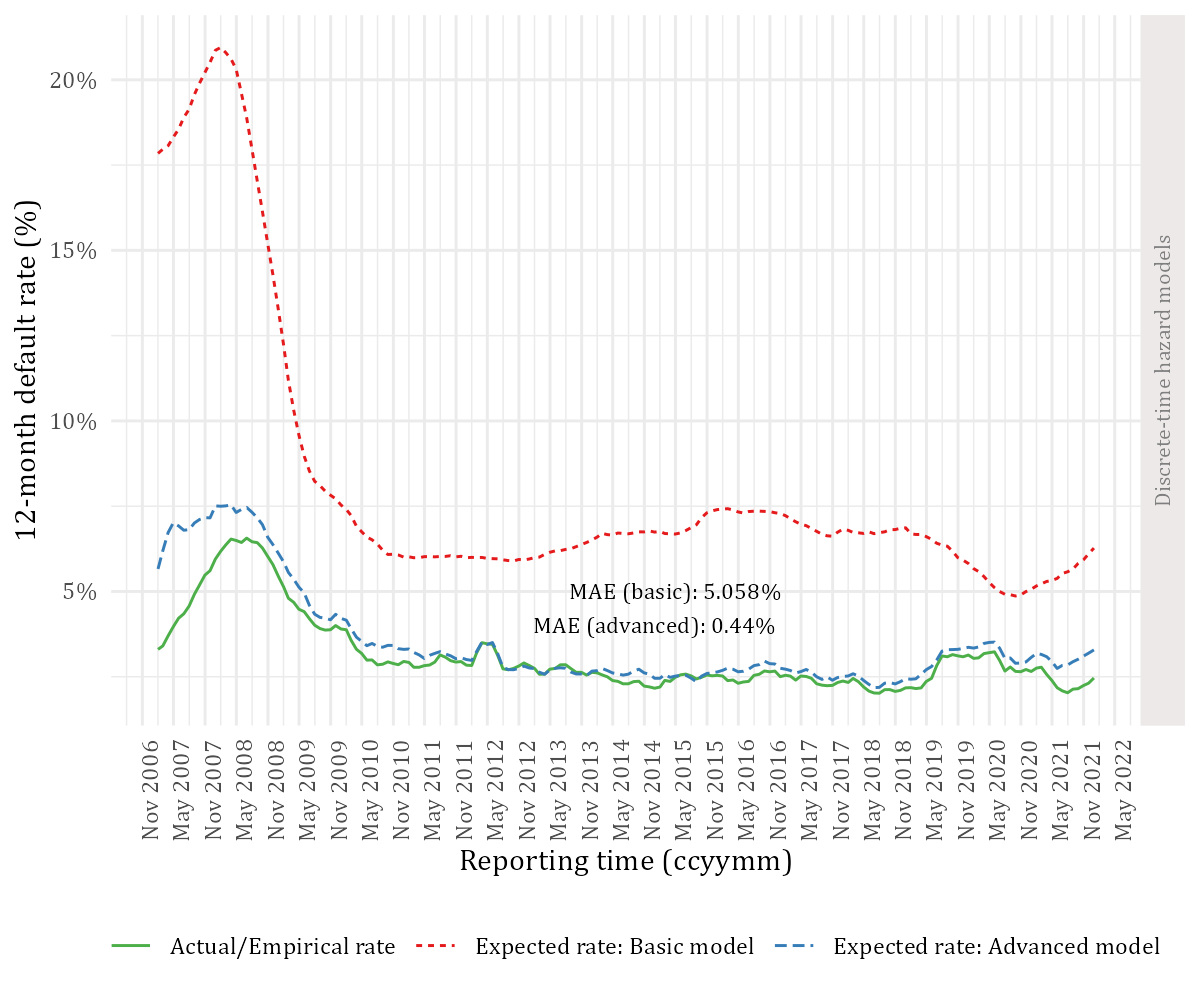}
    \caption{Showing the time graphs of 12-month default rates by model. The average discrepancy between the empirical rate and the expected rate per model is summarised using the MAE, as overlaid.}
    \label{fig:DefaultRates}
\end{figure}

One may also measure the time-dependent model calibration (as our fourth diagnostic) using a time graph of empirical vs expected 12-month default rates, which is called the \textit{rate comparison} diagnostic. Close agreement between such a pair of rates indicates better model performance in predicting real-world phenomena with greater accuracy. In particular, and at the loan level, the 12-month expected PD is calculated as the rolling sum of the 1-month event probabilities $f(t,\boldsymbol{x}_{ij})$ given input variables $\boldsymbol{x}_{ij}$ of subject-spell $(i,j)$ over its duration. Thereafter, one may calculate the mean of these predicted probabilities at each time point (or calendar month), which results in a time series that can be compared. This operation is programmed and graphed in script 6e of the R-codebase by \citet{botha2025termStructureSourcecode}, which culminates in \autoref{fig:DefaultRates}. The empirical 12-month default rate (green-solid) reacts as expected to the 2008 \textit{global financial crisis} (GFC), and is only moderately affected by the Covid-19 crisis. As before, the advanced DtH-model (blue-dashed) outperforms the basic DtH-model (red-dotted) in that the former model produces 12-month mean PD-values that are much closer to the empirical rates than those rates of the latter model. The calibration success is made measurable by again using the MAE between each pair of line graphs. In this instance, the advanced DtH-model scored an MAE (0.44\%) that is more than 11 times smaller than that of the basic DtH-model (5.058\%). This result is also visually corroborated in that the expected rates from the basic DtH-model far exceed the empirical rates across all time periods, particularly so during the 2008-GFC. Considering all of these results together, the superior model is clearly the advanced DtH-model, thereby underscoring the value of feature engineering, variable selection, and a weighted training scheme (i.e., WLR).

Consider now the three principles to which any term-structure model ought to adhere, as developed by \citet{skoglund2017}, and as previously discussed in \autoref{sec:background_reviews}. Firstly, we argue that the remarkably high tAUC-values of our DtH-models are testament to their ability to produce a monotonic risk-ordering, which agrees with the principle of risk-monotonicity. Secondly, the DtH-models are inherently conditioned on the economy since their input spaces contain macroeconomic variables by design. This macroeconomic sensitivity is also reflected in \autoref{fig:DefaultRates} in that the aggregated predictions of both DtH-models react as expected to known macroeconomic calamities, such as the 2008-GFC. Thirdly, the advanced DtH-model can replicate the observed term-structure and empirical 12-month default rate to a much greater degree than the basic DtH-model, which embodies the third principle of accuracy. It is therefore clear that the envisaged DtH-modelling approach obeys all three principles from \citet{skoglund2017}, which surely engenders additional assurance.

%% file: 3.6-Considerations.tex
\subsection{General considerations for applying the methodology across different product types}
\label{sec:Revolving}

Our analysis, data formulation, and modelling were conducted on a fixed-term product (residential mortgages), which might raise the question of applicability across other types of lending products.
However, we affirm that the underlying modelling methodology and data formulation (see \autoref{tab:dataStructure_perfSpells_PWP}) remains the same across other product types, e.g., revolving credit facilities or credit cards. 
Both of these examples can typically have longer lifespans than those of fixed-term products, particularly when the underlying portfolio is old (e.g., 30+ year credit card accounts). These longer lifespans would naturally result in performing spells that are longer in duration, depending on data. Moreover, longer lifespans imply greater opportunity for observing recurrent performing spells as an account defaults and cures multiple times. In our data formulation, the treatment of time, i.e., the spell period $t_{ij}$ for spell $j$ of loan $i$, as encoded in the field \texttt{TimeInPerfSpell}, remains the same across product type. This variable would however be equivalent to a "time on book" type of variable, though only for the first performing spell. Upon defaulting and curing, the $t_{ij}$-quantity naturally resets to 1 since a new performing spell has begun, while the spell number $j$ increments; itself encoded as the field \texttt{PerfSpell\_Num}. This dynamic between $t_{ij}$ and $j$ ensures that each row in the modelling dataset is conditionally independent from one another, given recurrent performing spells.

However, certain considerations are necessary when \textit{deploying} a DtH-model in general, depending on how the field \texttt{TimeInPerfSpell} is incorporated into the model towards embedding the baseline hazard. If a binning scheme is used, as in our case, then it is important that the last bin be open-ended to accommodate spells with durations beyond those observed in the training data. For example, a bin such as [300, 360] months may need to be adjusted to [300, max(\texttt{TimeInPerfSpell})] when deploying the DtH-model into production. Similar considerations would apply when incorporating \texttt{TimeInPerfSpell} via dummy variable encoding. Alternatively, if \texttt{TimeInPerfSpell} is used directly as an input variable, then no special considerations need apply for model deployment. However, and in our experience, directly using \texttt{TimeInPerfSpell} can lead to a poorer model fit and weaker discriminatory power. This is particularly true given the non-linearities in the baseline hazard over time, as alluded to in the non-linear empirical event probabilities in \autoref{fig:KaplanMeier_EventProb}. Therefore, a trade-off would appear to exist between ease of implementation and model performance. Nevertheless, and in summary, our modelling methodology and data formulation remain robust across product type, with only minor implementation adjustments needed for the way in which time is treated.

%% file: 4-Conclusion.tex
\section{Conclusion}
\label{sec:conclusion}

A variety of approaches exist for modelling the term-structure of default risk under IFRS 9, yet there is little consensus on the most correct approach, especially in retail lending. This lack of consensus is further exacerbated by the particular challenges faced in producing PD-estimates, which ought to be dynamic and time-dependent under IFRS 9.
Firstly, the default event can recur over the lifetime of a loan, which speaks to the dynamicity of credit risk. Secondly, a loan may be subject to other risks such as settlement or write-off, both of which preclude the main default event from occurring. Notwithstanding these two challenges (recurrent defaults \& competing risks), it remains notoriously difficult to produce a set of lifetime PD-estimates that are suitably dynamic and accurate, especially when trying to incorporate the broader macroeconomic environment. As such, we reviewed a few common approaches that are currently in vogue in modelling these lifetime PD-estimates, also known as the term-structure of default risk. This discussion naturally leads to listing the merits and limitations of each class of techniques, which had to be synthesised across various branches of literature. We believe that this review can itself serve as a guiding light to the practitioner in offering a menu of options for lifetime PD-modelling.

As our main contribution, we developed an in-depth and data-driven tutorial on modelling lifetime PD-estimates using discrete-time hazard (DtH) models, which is accompanied by an R-based codebase by \citet{botha2025termStructureSourcecode}. This step-by-step tutorial introduces standardised notation within the context of credit risk modelling, which is particularly useful for illustrating the way in which credit data ought to be prepared for survival analysis, i.e., recurrent performing subject-spells. 
Thereafter, a censoring study is illustrated towards motivating the use of survival analysis in the first place, followed by a histogram of failure times grouped by the type of spell resolution. Both graphs can serve diagnostic purposes in guiding the modelling strategy in practice, as demonstrated.
A resampling scheme is illustrated, which culminated in the development of a diagnostic measure (the \textit{resolution rate}) for evaluating the representativeness of a resampling scheme, thereby warding against sampling bias.
Following this, we formulated and demonstrated the empirical term-structure of default risk as the collection of discrete event probabilities (or marginal PDs) over time, having used the framework of survival analysis. In approximating these probabilities, two competing DtH-models (basic vs advanced) are fit using a varied set of input variables, including macroeconomic variables. These models are extensively evaluated using four diagnostic measures, some of which are newly formulated within the context of recurrent subject-spells. These diagnostics include: 1) the \textit{time-dependent area under the curve} (tAUC) for evaluating discriminatory power; 2) the \textit{time-dependent Brier score} (tBS) for assessing prediction accuracy and calibration; 3) the empirical vs expected term-structures; and 4) the time graphs of 12-month empirical vs expected default rates. Both of the last two diagnostics contend with the calibration of a model to reality. Our results show that the advanced DtH-model massively outperforms the basic DtH-model across most diagnostic measures, which is testament to proper feature engineering and modelling practices. Lastly, our set of input variables should themselves shed light on the type of risk drivers that proved instrumental in the success of the advanced DtH-model.

Ultimately, we believe that this tutorial can become crucial in cultivating best practices when using survival analysis in modelling lifetime PD-estimates under IFRS 9. These best practices should be of great value to model validators, practitioners, and regulators alike.
Moreover, an unexpected benefit is that the tutorial employs logistic regression (LR) in fitting DtH-models, which is a technique that already enjoys ubiquity in banking. Accordingly, we believe that the fitting of DtH-models is more accessible than previously thought.
However, and given the nature of any tutorial, some aspects can certainly be improved in future work. In the interest of expediency, we followed the latent risks approach in contending with competing risks, which are duly marked as right-censored. Future researchers can rather develop a suite of competing risk survival models, perhaps using the Fine-Gray approach from \citet{putter2007tutorial}, which is the more correct approach to dealing with competing risks.

Furthermore, our DtH-models are fit to the entirety of a portfolio, without using any segmentation and relying solely on the quality of input variables to differentiate risk. Conversely, segmentation may further improve model performance in some aspects, provided that sufficient data exists within each partition. However, more partitions imply having more models to fit and to maintain, which can invite unnecessary model risk when compared to a single model with a more sophisticated input space. The prudent practitioner therefore guards against over-zealously partitioning the data.
Moreover, while we focused on using a logit link function within a GLM-framework, other options are certainly possible. In particular, future work could explore machine learning varieties of survival analysis such as random survival forest and ensemble techniques.
Lastly, some of the input variables within our DtH-models are dynamic over time (or `behavioural'). These behavioural variables would need forecasting when implementing a DtH-model for prediction purposes. While this forecasting is not typically a problem for macroeconomic variables (since many banks already produce such forecasts themselves), it might be problematic for other behavioural variables. Future research can certainly explore the viability of such forecasting mechanisms, which may include the naïve method (last observed value), or simply fitting the variable as a function of spell time.

%% file: 5-Branch_PD-Reuse.tex
\section{Appendix: Approaches that re-use existing PD-models in deriving lifetime PDs}
\label{app:PD_reuse}

In \autoref{sec:Basel_GLM}, we review and supplement a simple framework from \citet[\S 3.2]{bellini2019} that re-uses Basel-based PD-estimates by adjusting them with macroeconomic forecasts using a GLM in producing lifetime PD-estimates. There exist two varieties of this approach: a portfolio-level version for when data is limited, and an account-level version that is more viable though requires more granular data. Additionally, some high-level principles are summarised regarding the validation of models arising from this Bellini-approach.
Another approach that re-uses PD-estimates is that of \citet{breed2021}, which we summarise and demonstrate in \autoref{sec:empiricalTermStructure}.
Having synthesised all approaches using standardised notation, we critically review each method and outline their merits and drawbacks.

\input{5.1-Bellini2019}
\input{5.2-Breedt}

%% file: 5.1-Bellini2019.tex
\subsection{The Basel-based GLM-approach from \citet{bellini2019}}
\label{sec:Basel_GLM}

The first approach from \citet[\S3.2]{bellini2019} is based on a GLM-framework that reuses the one-year PD-estimates from the Basel-context in deriving the term-structure under IFRS 9. In particular, consider the Bernoulli random variable $Y\in\{0,1\}$ where 1 flags a default and 0 indicates a non-default event over a one-year outcome period. Let $y_{it}\in Y$ be historical default realisations for borrowers $i=1,\dots,N_p$ over discretely-measured calendar time $t=t_1,\dots,t_n$. Let $\mathcal{S}_\mathrm{P}(t)$ represent a set of performing (or non-defaulted) accounts at $t$ that are subject to default risk, and let $n'_t$ signify the volume of accounts in $\mathcal{S}_\mathrm{P}(t)$. Consider then the portfolio-level 12-month conditional default probability $\mathbb{P}(Y_{t+12}=1 \, | \, Y_t =0) $, where $Y_t, Y_{t+1},\dots$ are Bernoulli random variables over $t$. In following the \textit{worst-ever} aggregation approach from \citet[\S 3.1.3]{botha2021phd}, one can estimate a variant of this probability at each $t$ within $\mathcal{S}_\mathrm{P}(t)$ with the \textit{default rate}, which is itself defined as the time series over $t$ as
\begin{equation} \label{eq:defaultRate}
    D_t = \frac{1}{n'_{t}}\sum_{i\, \in \, \mathcal{S}_\mathrm{P}(t)}{ \mathbb{I}\left(\max{(y_{it}, \dots, y_{i(t+12)})=1} \right)} \, ,
\end{equation}
where $\mathbb{I}(\cdot)$ is an indicator function. This time series may also be estimated within subregions of the portfolio, or `sub-portfolios', in following a segmentation approach; though we shall restrict ourselves to the entire portfolio in the interest of simplicity.

In what \citet[\S3.3.1]{bellini2019} called the \textit{portfolio-level sub-approach}, the time series $D_t$ from \autoref{eq:defaultRate} forms an overall `creditworthiness' index over time and may be regressed upon historical realisations from $p$ macroeconomic variables (MVs), denoted by $\boldsymbol{z}_t=\left\{z_1,\dots,z_j,\dots,z_p \right\}$. In particular, the default rates can be modelled as $g(\mu_t)=\eta_t$ with an appropriate link function $g$, where $\mu_t=g^{-1}(\eta_t)=\mathbb{E}(Y_t \, | \, \boldsymbol{z}_t)$ is the mean default rate at time $t$, and $\eta_t$ is a linear predictor defined as
\begin{equation} \label{eq:LP_time}
    \eta_{t} = \beta_0 + \beta_1z_1 +\dots \beta_pz_p \, ,
\end{equation}
where $\boldsymbol{\beta}=\left\{\beta_0, \beta_1, \dots, \beta_p \right\}$ are estimable regression coefficients. 
\citet[\S3.3.1]{bellini2019} provided an example hereof based on the identity link $g(\mu_t)=\eta_t=\mu_t$, having fit the historical default rate as a function of a few MVs. Then, the predictions $\hat{\mu}_t$ given forecasts $\hat{\boldsymbol{z}}_t$ over the time horizon $t=t_{n+1},\dots,t_{n+h}$ are funnelled through an additional "shifting" function $f_s$. The simplest variant\footnote{\citet[pp.149--151]{bellini2019} also provided a rather convoluted shifting function called the "logit shift", which introduces elements of account-level modelling into an approach originally described as `portfolio-level'. However, and under such an approach, account-level data is typically unavailable and we shall therefore not explore this logit shift function any further.} hereof is defined as the so-called "proportional shift", given as 
\begin{equation} \label{eq:propShift}
    f_\mathrm{s}\left(x, \hat{\mu}_t \right) = x \frac{\hat{\mu}_t}{ \hat{\mu}_{t_n}} \, .
\end{equation}

In the penultimate step of this approach from \citet[\S3.3.1]{bellini2019}, we consider the initially-available PD-estimate $p(\boldsymbol{x}_i)$ from the Basel-context given input variables $\boldsymbol{x}_i=\left\{x_1,\dots\right\}$ of borrower $i$ at the last observed time $t_n$. This PD-estimate, which typically contains zero MVs, is then attenuated to macroeconomic forecasts using \autoref{eq:propShift} and duly updated as the "macro-shifted PD", defined over $t$ as
\begin{equation} \label{eq:PD_macro}
    \dot{p}(\boldsymbol{x}_i, \hat{\mu}_t) = f_\mathrm{s}\left(p(\boldsymbol{x}_i), \hat{\mu}_t \right)  \, .
\end{equation}
A rudimentary survival probability $S(t,\boldsymbol{x}_i)$ up to time $t$ and given $\boldsymbol{x}$ is then derived using a series of these macro-shifted PD-estimates from \autoref{eq:PD_macro}, which are themselves treated as discrete hazards within a survival analysis context. More formally, this survival probability is expressed using $\dot{p}(\boldsymbol{x}_i, \hat{\mu}_t)$ up to the forecast time $t=t_{n+1},\dots,t_{n+j}\dots,t_{n+h}$ as
\begin{equation} \label{eq:survival_basel}
    S(t, \boldsymbol{x}_i) = \prod_{q=t_n}^{t}{\left( 1-\dot{p}\left(\boldsymbol{x}_i, \hat{\mu}_{q}\right) \right)} \, .
\end{equation}
Finally, the lifetime PD-estimate is then estimated by multiplying $S(t-1, \boldsymbol{x}_i)$ from \autoref{eq:survival_basel} with $\dot{p}(\boldsymbol{x}_i, \hat{\mu}_t)$. Doing so resembles the well-known identity in discrete-time survival analysis of the probability mass function $f(t)=S(t-1)\cdot h(t)$, where $h(t)=\dot{p}(\boldsymbol{x}_i, \hat{\mu}_t)$ is the hazard function. As such, the lifetime PD-estimate $\hat{p}(\boldsymbol{x}_i,t)$ is given at any particular forecast time $t\geq t_{n+1}$ by
\begin{equation} \label{eq:PD_lifetime_basel}
    \hat{p}(\boldsymbol{x}_i,t) = S(t-1,\boldsymbol{x}_i) \cdot \dot{p}(\boldsymbol{x}_i, \hat{\mu}_t) \, .
\end{equation}

Whilst certainly modular and simple, the Basel-based portfolio-level GLM-approach from \citet[\S3.3.1]{bellini2019}, or the self-styled "Bellini-approach", has its drawbacks. Most notably, an inappropriate choice of the link function $g$ can produce an output $g(\mu_t)$ that is no longer a probability, thereby misrepresenting the underlying creditworthiness index; i.e., the default rate $D_t$.
Furthermore, the way in which the PD-estimate is adjusted in \autoref{eq:propShift} assumes that macroeconomic effects are equally distributed across all loans; which is quite a strong assumption. It may very well be that certain segments of higher risk-grades are more acutely affected by certain macroeconomic effects than lower risk-grades.
Moreover, assume that the PD-estimate $p(\boldsymbol{x}_i)=\hat{\mu}_i$ arises itself from another GLM (with a logit link) in estimating the account-level default probability $\mathbb{P}(Y=1 \, | \, \boldsymbol{x}_i)$. Adjusting $\hat{\mu}_i$ in such a \textit{post hoc} and exogenous manner as in \autoref{eq:PD_macro} will compromise the estimation process of $\boldsymbol{\beta}$, which may in turn bias the estimate of the mean default probability $\mu_i=\mathbb{E}(Y\,|\, \boldsymbol{x}_i)$. This bias would violate the third principle (Accuracy) from \citet{skoglund2017} in that the eventual lifetime PD-estimates may no longer replicate the observed experience as accurately as possible; even though they are `conditioned' on the economy.
Lastly, it is assumed that $\dot{p}(\boldsymbol{x}_i, \hat{\mu}_t)$ operates like a hazard rate. However, its estimation does not at all guarantee that the output adheres to it being the failure probability during the discrete interval $(t-1,t]$, given survival up to $t-1$. Overall, trying to re-use Basel-based PD-models is likely to be suboptimal and we agree with the sentiment of \citet{bank2021review}, who described such re-use as a rather "coarse implementation of IFRS 9". 
For these reasons, we deem the Bellini-approach to be less suitable for an IFRS 9 modelling context, except perhaps when developing a baseline model.

In countering some of these limitations, \citet[\S3.3.2]{bellini2019} also provided an \textit{account-level sub-approach}, which foregoes regressing the portfolio-level default rate $D_t$ from \autoref{eq:defaultRate} and rather focuses on (re)-modelling the PD at the account-level. Doing so embeds the MVs $\boldsymbol{z}_t$ endogenously into the PD-model as a part of the overall input space $\boldsymbol{x}=\left\{ \boldsymbol{z}_t,\boldsymbol{x}_i \right\}$. In this case, the historical default realisations $y_{it}\in Y$ are regressed upon the variables in $\boldsymbol{x}$ in modelling the account-level default probability $\mathbb{P}(Y=1\, | \, \boldsymbol{x})$. From \citet[pp.~1-10]{hosmer2000logistic}, a logit link function $g$ is commonly used in modelling the mean default probability $\mu_{it}$ of borrower $i$ at time $t$. More formally, we define this logit link function as
\begin{equation}
    g(\mu_{it})=\log{\left( \frac{\mu_{it}}{1-\mu_{it}} \right)} = \mathrm{logit}(\mu_{it}) \quad \implies \quad \mu_{it} = g^{-1}(\eta_{it})=\left(1+ e^{-\eta_{it}} \right)^{-1} \, ,
\end{equation}
where $\eta_{it}$ is the linear predictor across the (redefined) input space $\boldsymbol{x}$ with estimable regression coefficients $\boldsymbol{\beta}$, as in \autoref{eq:LP_time}.
Given forecasts $\hat{\boldsymbol{x}}$ of the input space, the author continues by macro-shifting the prediction $\hat{\mu}_{it}$ using the proportional shift function from \autoref{eq:propShift}. However, the reason for doing so (to embed MVs further) is unconvincing, particularly since $\hat{\mu}_{it}$ already contains macroeconomic information by design.
Finally, the process by which lifetime PD-estimates are obtained is exactly the same as for the previous portfolio-level sub-approach, i.e., applying \autoref{eq:PD_lifetime_basel} over $t$ given the macro-shifted version of $\hat{\mu}_{it}$. Despite its relative simplicity, this approach therefore attracts many of the same criticisms of the previous variant.

\citet[\S 3.3.3]{bellini2019} outlined a few principles for validating the aforementioned lifetime PD-estimates, and briefly demonstrated two of these principles. 
Firstly, the author examined the discriminatory power of these estimates using a classical ROC-analysis, as summarised by the AUC. Greater AUC-values indicate greater discriminatory power. 
Secondly, the calibration success of a model is briefly described, which speaks to the accuracy of the predictions produced from the fitted model. The author illustrated such an exercise by comparing the portfolio-level default rate to the aggregated PD-estimates at each time point. Closer agreement between either time series (actual vs expected) is desired, and was measured using a correlation study.
Some other validation principles include the following. The modelling method should be verified in accordance with the available data. The model parameters ought to be analysed, as should be the stability of the model within a data sample that was not used during model training. Marginal (or variable importance) analysis can help elucidate the degree to which each variable contributes to model predictions. Lastly, one should assess the extent to which the model can be rebuilt independently, which usually requires comprehensive documentation to exist.
While these principles are certainly plausible, they were not extensively demonstrated by the author; an aspect upon which our work can improve.

%% file: 5.2-Breedt.tex
\subsection{The empirical term-structures approach from \citet{breed2021}}
\label{sec:empiricalTermStructure}

\citet{breed2021} proposed a methodology for deriving a set of PD-estimates under IFRS 9, which also culminates in a term-structure of default risk. This method has its genesis in a so-called \textit{defaults table}, shown in \autoref{tab:exampledefaults}, which contains the number of observed defaults over the lifetime of each cohort.
As in \autoref{sec:Basel_GLM}, let $y_{itv}\in Y$ be historical default realisations for borrowers $i=1,\dots,N_p$ in cohort $t$ over discretely-measured time points $v=1,2,\dots$ of the particular cohort's lifetime. Furthermore, let ${S}_{\mathrm{P}}$ represent a set of at-risk accounts in cohort $t$ that were performing as at time $v$. From this defaults table, the empirical one-month default rate for cohort $t$ at point $v$ of its lifetime is calculated as
\begin{equation} \label{eq:EmpDefault}
    D_{tv}^E = \frac{1}{n'_{tv}}\sum_{i\, \in \, \mathcal{S}_\mathrm{P}(t,v)}{ y_{itv}} \, = \frac{d_{tv}}{n'_{tv}},
\end{equation} 
where $n'_{tv}$ is the number of accounts that survived during the interval $[v-1,v]$, $n'_{t0}$ is the initial account volume of cohort $t$, and $d_{tv}$ is simply the number of defaults in cohort $t$ at point $v$. 

\begin{table}[htbp]
\centering
\caption{A defaults table showing the number $d_{tv}$ of defaults across cohorts $t=t_1,\dots,t_M$ over their lifetimes $v=1,2,\dots$. From \citet{breed2021}.}
\label{tab:exampledefaults}
\begin{tabular}{@{}llccccccc@{}}
\toprule
\multirow{2}[3]{*}{
\textbf{Cohort} ($t$)} & \multirow{2}[3]{*}{\textbf{Initial account volume} ($n'_{t0}$)} & \multicolumn{7}{c}{\textbf{Lifetime $(v)$} } \\
\cmidrule(l){3-9}
& & 1 & 2 & 3 & 4 & 5 & 6 & 7 \\
\midrule
201501 & 500 & 10 & 5 & 4 & 8 & 6 & 3 & 3 \\
201502 & 550 & 11 & 5 & 6 & 3 & 7 & 5 &    \\
201503 & 600 & 13 & 5 & 7 & 4 & 6 &   &    \\
201504 & 650 & 14 & 6 & 6 & 5 &   &   &    \\
201505 & 700 & 15 & 5 & 7 &   &   &   &    \\
201506 & 750 & 14 & 7 &   &   &   &   &    \\
201507 & 800 & 16 &   &   &   &   &   &    \\
\bottomrule
\end{tabular}
\end{table}

\citet{breed2021} needed to aggregate across the various cohorts in deriving a term-structure of values over $v$. They did so by calculating the volume-weighted average default rate over a so-called \textit{reference period}, which has length $r$ (in months). This operation is anchored at the last cohort $t_M$ and sequentially executed backwards on each previous cohort, having fixed the time point (or horizon) $v$ to a specific value. In so doing, the horizon-specific empirical\footnote{\citet{breed2021} called this quantity the `marginal PD', yet doing so in our work would conflict with the definition of $f(t)=S(t-1)-S(t)$ as the discrete density (or marginal probability) of the main event, at least within the context of survival analysis.} PD is calculated over each interval $[\tau-r,\tau]$ for cohorts $\tau=t_M,t_{M-1},\dots,t_r$, thereby aggregating defaults across cohorts. We repeat this operation for successive $v$-values, and formalise the estimation of the empirical lifetime PD over each horizon $v$, given a fixed reference period $r$, as
\begin{equation} \label{eq:PD_breed}
    \tilde{p}(v,r) = \frac{\sum_{w=\tau-(v-1)-(r-1)}^{\tau-(v-1)}{d_{wv}}}{\sum_{w=\tau-(v-1)-(r-1)}^{\tau-(v-1)}{n'_{w0}}} \quad \text{for} \ \tau=t_M,t_{M-1},\dots,t_{r} \, .
\end{equation}
By way of an example, consider setting $r=3$ months, whereupon we calculate the empirical PDs, starting from the last three cohorts 201505 to 201507. For $v=1$, the corresponding number of total defaults ($15 + 14 + 16=45$) is divided by the number of at-risk accounts ($700+750+800=2250$), yielding an empirical PD of 2\%.
Repeating this operation across all $v$-values results in an empirical term-structure of default risk, as shown in \autoref{tab:pd_term_structure}.

\begin{table}[htbp]
\centering
\caption{An example of calculating the empirical PD term-structure across horizon $v$ for a reference period of $r=3$ months. From \citet{breed2021}.}
\label{tab:pd_term_structure}
\begin{tabular}{@{}ccccc@{}}
\toprule
\textbf{Horizon} & \textbf{Cohorts} & \textbf{Sum of Performing Accounts} & \textbf{Sum of Defaults} & \textbf{Empirical PD} \\
\midrule
1 & 201505 : 201507 & 2250 & 45 & 2.000\% \\
2 & 201504 : 201506 & 2100 & 18 & 0.857\% \\
3 & 201503 : 201505 & 1950 & 20 & 1.026\% \\
4 & 201502 : 201504 & 1800 & 12 & 0.667\% \\
5 & 201501 : 201503 & 1650 & 19 & 1.152\% \\
\bottomrule
\end{tabular}
\end{table}

The described methodology will result in a single term-structure, which would not be sensitive to the characteristics $\boldsymbol{x}_i$ of the individual loan account $i$. As a possible remedy, \citet{breed2021} posited that one may partition the data into a series of non-overlapping segments, each of which captures some homogeneous aspect of borrower behaviour. \autoref{eq:PD_breed} is then calculated within each segment $\mathcal{S}$, whereby $\tilde{p}(v,r)$ is assigned to each member account $i\in\mathcal{S}$. The goal of such segmentation is to derive PD term-structures that are structurally different from one another towards ensuring appropriate risk differentiation. 
\citet{breed2021} also discussed another approach by which account-level term-structures can be derived, which is called Lorenz curve estimation. As explained in-depth by \citet{globner2003basel}, this process leverages behavioural credit scores $s=f(\boldsymbol{x}_i)$, whose constitution is assumed to encompass many account-level input variables. This approach ultimately results in a scalar that is multiplied with the horizon-specific empirical PD-estimate from \autoref{eq:PD_breed}, thereby risk-sensitising $\tilde{p}(v,r)$ to the variables $\boldsymbol{x}_i$ of a particular account $i$ in predicting its default risk.

Whilst certainly simple and original, the self-styled Breed-method is not without its limitations. 
Firstly, and in risk-sensitising the term-structure, the number and composition of segments will greatly influence the results. Too many segments (or too fine a granularity) may fail to avail a sufficient number of data points, whereas too few segments may no longer differentiate default risk across the portfolio. 
Secondly, an inappropriately small value for the reference period $r$ may introduce seasonal effects into the term-structure at certain time points $v$. Conversely, overly large $r$-values can result in term-structures that are inadequate in length, e.g., less than 12 months.
Thirdly, and considering the Lorenz curve calibration, the adjustment of $\tilde{p}(v,r)$ using account-level scalars can compromise the original estimation process of $\tilde{p}(v,r)$ and most certainly affect its calibration to observed default experience. This is because this scalar-based approach effectively weighs $d_{tv}$ in \autoref{eq:PD_breed} differently for each credit score $s$. As with the GLM-based scalar within the Bellini-approach in \autoref{sec:Basel_GLM}, this process can introduce unintended model risk and even bias into the lifetime PD-estimates, which would again detract from the third principle (Accuracy) of \citet{skoglund2017} for deriving term-structures. This is especially true for Lorenz curve calibration since the underlying model (a behavioural scorecard) is not necessarily fit-for-purpose under IFRS 9, at least compared to a more direct and bespoke modelling approach such as survival analysis.
Fourthly, the success of following a Lorenz curve calibration is directly dependent on the discriminatory power of another model, which implies that any issues with said model can bleed into the eventual PD term-structure itself.

There are other ways of risk-sensitising the empirical term-structure from the Breed-method, especially since the behavioural credit score $s$ does not typically include macroeconomic variables (MVs) in its constitution, as explored by \citet{ brunel2016lifetime}.
In particular, \citet{Breed2023} proposed a framework that is based on \textit{principal component regression} (PCR) in adjusting a lifetime PD term-structure given by \autoref{eq:PD_breed} using a "macroeconomic scalar".
Leveraging the historical default rates $D_t$ from \autoref{eq:defaultRate}, the authors first derived a \textit{credit risk index} (CRI), i.e., the rolling average default rate over the last 12 monthly cohorts of at-risk accounts. This CRI represents the overall behaviour of a loan portfolio over time, and is ultimately regressed upon a collection of MVs (including lags thereof), as constituted using principal component analysis. The authors favourably compared their PCR-model to a few other techniques in directly modelling the CRI as a function of MVs. In each case, the resulting model is fed scenario-based forecasts of MVs, thereby producing a macroeconomic scalar.
This concept is further developed by \citet{Moodley2025}, who also advocated for integrating such a macroeconomic scalar into expected credit loss models. They developed practical guidance towards adjusting existing PD-estimates using macroeconomic information, and demonstrated the method using datasets from Kenya and Mauritius. This method is particularly apt where MVs are limited and volatile, which are typical features of developing economies.
One should however remain cognizant of it limitations, particularly since this scalar-based method assumes all borrowers across all risk grades will be equally affected by a macroeconomic change; which is unlikely to hold in reality.


%% file: 6-Diagnostics.tex
\section{Appendix: The formulation of time-dependent diagnostics for survival models}
\label{app:tDiagnostics}

In \autoref{sec:tROC}, we provide a succinct summary of time-dependent ROC-analysis in evaluating the discriminatory power of a survival model over time. Thereafter, a high-level review follows in \autoref{sec:tBS} of time-dependent Brier scores, which are useful in measuring the prediction accuracy of a survival model over time. Both of these works are specifically formulated within the context of credit risk modelling, i.e., recurrent subject-spells.

\input{6.1-tROC}

\input{6.2-tBS}

%% file: 6.1-tROC.tex
\subsection{Using time-dependent ROC-analysis (tROC) for evaluating discriminatory power}
\label{sec:tROC}

Every binary classifier model has some measurable ability in distinguishing those observations that are at greater risk of the main event from those that are at lower risk. The assessment of a model's discriminatory power typically culminates in a classical \textit{receiver operating characteristic} (ROC) curve, which plots the trade-off between the true positive rate ($T^+$) and false positive rate ($F^+$); see \citet{fawcett2006introduction}. However, \citet{botha2025recurrentEvents} explained that certain outcomes will lack an outcome depending on the chosen time frame over which assessments are made, particularly so in survival analysis. The classical ROC-curve assumes that all observations are fully observed against which predictions are then tested, which implies that it would be inadequate in measuring the discriminatory power of a survival model amidst right-censored cases.
In addressing this issue, \citet{heagerty2000} and \citet{bansal2018tutorial} generalised the ROC-framework by defining $T^+$ and $F^+$ as time-dependent quantities. Let \( M \) be a random variable that denotes the risk scores (or marker values) emanating from a survival model given covariates, such that greater $M$-values represent greater risk of the event, and vice versa. Consider $T$ as a non-negative and discrete-valued random variable that denotes latent lifetimes. Then, let $T_{ij} \in T$ be the observed event time (or spell age) of subject-spell $(i,j)$, as defined in \autoref{eq:timeSpent_performing}, where $i = 1,\dots,N_p$ indexes loans, and $j = 1,\dots,n_i$ indexes the performing spells per loan $i$. Define \( D_{ij}(t) = \mathbb{I}(T_{ij} \leq t) \) as the event indicator at time $t$ of the spell, where $\mathbb{I}(\cdot)$ is an indicator function. Then, $m_{ijt}\in M$ are realisations of $M$ over the spell period $t=1,\dots,T_{ij}$. This notation forms the basis from which we will present the time-dependent ROC-framework.

Binary event predictions are rendered by dichotomising the marker values $m_{ijt}$ using a threshold $p_c$, i.e., predicting an event if $m_{ijt} > p_c$, and a non-event otherwise. The time-dependent true and false positive rates under the cumulative case/dynamic control (CD) framework, itself explained by \citet{bansal2018tutorial}, are then given for a particular $p_c$-value as
\begin{align} \label{eq:TPR}
    T^+(p_c,t) &= \mathbb{P}(M > p_c \mid T \leq t) = \mathbb{P}(M > p_c \mid D_{ij}(t) = 1) \, , \\
    F^+(p_c,t) &= \mathbb{P}(M > p_c \mid T > t) = \mathbb{P}(M > p_c \mid D_{ij}(t) = 0) \, . \label{eq:FPR}
\end{align}
In following \citet{heagerty2000}, we can use Bayes' theorem together with the conditional survivor functions $S(t \mid M > p_c)$ and $S(t \mid M \leq p_c)$, themselves estimated only within the subsets of those cases with markers $m_{ijt}>p_c$ and $m_{ijt}\leq p_c$ respectively. Then, the expressions from \crefrange{eq:TPR}{eq:FPR} become
\begin{align} \label{eq:tpr_bayes}
    T^+(p_c,t) &= \frac{(1 - S(t \mid M > p_c)) \, \mathbb{P}(M > p_c)}{1 - S(t)}, \\
    F^+(p_c,t) &= 1 - \frac{S(t \mid M \leq p_c) \, \mathbb{P}(M \leq p_c)}{S(t)} \label{eq:fpr_bayes}.
\end{align}

To estimate these quantities, one can use the \textit{Nearest Neighbour} (NN) estimator from \citet{akritas1994}, as formulated by \citet{heagerty2000}. The authors used this NN-estimator in calculating the bivariate survivor function \( S(p_c, t) = \mathbb{P}(M > p_c, T > t) \). However, the NN-estimator cannot readily contend with a multi-spell setup with time-varying covariates wherein observations are commonly clustered around a single subject-spell. As a remedy, \citet{botha2025recurrentEvents} proposed an adjustment that can work within our setup, which is defined as the 
\begin{equation}
\label{eq:S_NN_estimator_meanAdj}
    \text{mean-adjusted Akritas-estimator:} \quad 
    \hat{S}_{\lambda_n} (p_c,t) = \frac{1}{n} \sum_{(i,j)} \left\{
    \frac{1}{\eta_{ij}}\sum_{v=1}^{T_{ij}} 
    \hat{S}_{\lambda_n} \big(t \, \rvert \,M = m_v \big) \mathbbm{I} \big(m_{v} > p_c \big) \right\} \, .
\end{equation}
In \autoref{eq:S_NN_estimator_meanAdj}, \( \lambda_n \) is a smoothing parameter (as discussed later), $n$ is the number of subject-spells, and $\eta_{ij}$ is the size of the risk set that contains those qualifying marker values $m_v$, i.e., those markers where $m_{ijt}>p_c$ over $t_{ij}=1,\dots,T_{ij}$. Overall, \autoref{eq:S_NN_estimator_meanAdj} represents the average (conditional) survivor function for those markers that exceed $p_c$. 
In defining the \textit{marker-conditional} survivor function $S(t \mid M=m_v)$ within \autoref{eq:S_NN_estimator_meanAdj}, consider the default resolution indicator $\delta_{ij}\in\{0,1\}$, which equals 1 if the default/main event occurred for $(i,j)$ at some point, and 0 otherwise. Consider the sequence of all time-ordered markers $m_1,\dots,m_w,\dots,m_W$, where $W$ is the total number of markers, and the ordering is based on the subject-spell ages $T_{(1)}<\cdots <T_{(w)}< \cdots<T_{(W)}$. Given a marker $m_w$, we calculate $S(t \mid M=m_w)$ using the NN-related kernel function $K_{\lambda_n}\in\{0,1\}$ as a weight within a Kaplan-Meier type estimator, expressed as
\begin{equation}
\label{eq:S_NNE_weightedKM}
    \hat{S}_{\lambda_n}(t \mid M = m_w) = \prod_{q \leq t} \left\{ 1 - \frac{\sum_{s=1}^W K_{\lambda_n}(m_{s}, m_w) \, \mathbb{I}(T_{ij} = q, \delta_{ij}=1)}{\sum_{s=1}^W K_{\lambda_n}(m_{s}, m_w) \, \mathbb{I}(T_{ij} \geq q)} \right\},
\end{equation}

\citet{akritas1994} originally formulated the smoothing weights in \autoref{eq:S_NNE_weightedKM} using a 0/1 nearest neighbour kernel function $K_{\lambda_{n}}(m_s,m_w)$ across any ordered pair of marker values $(m_s,m_w)$ for $s\ne w$. In particular, 
\begin{equation} \label{eq:NN_kernel}
    K_{\lambda_n}(m_{s}, m_w) = \mathbb{I}\left( -v(\lambda_n) < \hat{F}_M(m_{s}) - \hat{F}_M(m_w) < v(\lambda_n) \right) \, ,
\end{equation}
where \( \hat{F}_M(\cdot) \) is the empirical marker distribution and \( 2\lambda_n \in(0,1)\) is the bandwidth, i.e., the proportion of observations included within each neighbourhood. Each neighbourhood is bounded by $\left[ -v(\lambda_n),v(\lambda_n) \right]$, where $v(\cdot)$ produces a neighbourhood bound from the underlying sequence of time-ordered markers; see \citet{heagerty2000}. Furthermore, \citet{botha2025recurrentEvents} proposed another mean-based adjustment to the empirical marker distribution in dealing with clustered observations, given as
\begin{equation} \label{eq:marker_CDF_meanAdj}
    \hat{F}_M(m) = \frac{1}{n} \sum_{(i,j)} \left\{ \frac{1}{\eta_{ij}} \sum_{v=1}^{T_{ij}} \mathbb{I}(m_v \leq m) \right\} \, .
\end{equation}
Finally, 
the time-dependent true and false positive rates from \crefrange{eq:tpr_bayes}{eq:fpr_bayes} are duly estimated as
\begin{align} \label{eq:tpr_NNE}
    T^+(p_c,t) &= \frac{(1 - \hat{F}_M(p_c)) - \hat{S}_{\lambda_n}(p_c,t)}{1 - \hat{S}_{\lambda_n}(t)}, \\
    F^+(p_c,t) &= \frac{\hat{S}_{\lambda_n}(p_c,t)}{\hat{S}_{\lambda_n}(t)} \, ,
\end{align}
where \( \hat{S}_{\lambda_n}(t) = \hat{S}_{\lambda_n}(p_c,t) \) for \( p_c = -\infty \).
\citet{botha2025recurrentEvents} refer to the use of \crefrange{eq:S_NN_estimator_meanAdj}{eq:marker_CDF_meanAdj} as the ``clustered tROC extension'' of time-dependent ROC analysis. We also provide an R-based implementation hereof via the \texttt{tROC.multi()} function in the codebase of \citet{botha2025termStructureSourcecode}, as compiled in script 0d and applied in script 6a on the basic and advanced DtH-models.

%% file: 6.2-tBS.tex
\subsection{Towards measuring calibration and discriminatory power: the time-dependent Brier score}
\label{sec:tBS}

From \citet{Graf1999}, it is difficult to predict the time-to-event, and effort should rather be spent on estimating at $t=0$ the probability that the default event will \textit{not} occur until a given time frame $t^*$, i.e., the survival probability.
The \textit{time-dependent Brier score} (tBS) considers this survival probability as a prediction, and generalises the traditional Brier score to time-to-event data with censoring.
Consider $T$ as a discrete non-negative random variable that denotes the latent lifetimes of subject-spells $(i,j)$, with $T_{ij}$ being realisations of $T$. Let $Y_{ij}(t^*)=\mathbb{I}(T_{ij}>t^*) \in\{0,1\}$ be the event-free status at a fixed time point $t^*$. Now let $\hat{p}_{ij}(t^*)\in[0,1]$ be the corresponding prediction thereof, i.e., the estimate of the conditional survival (or event-free) probability $\mathbb{P}\left(T > t^* \,| \, \boldsymbol{x}_{ij}\right)$ given covariates $\boldsymbol{x}_{ij}$. The tBS measures the mean squared differences between the predicted and observed event-free statuses, and is defined at $t^*$ as
\begin{equation} \label{eq:tBS}
    \mathrm{tBS}(t^*) = \mathbb{E}\left[ \left( Y_{ij}(t^*) - \hat{p}_{ij}(t^*) \right)^2 \right],
\end{equation}

Lower values of $\mathrm{tBS}(t^*)$ indicate better predictive accuracy, while $\mathrm{tBS}(t^*) = 0$ corresponds to perfect prediction at $t^*$. 
Regarding its estimation, \citet{Graf1999} and \citet{suresh2022survival} noted that one must account for the loss of information in $Y_{ij}(t^*)$ due to censoring. 
For each $(i,j)$, we therefore observe the censoring time $C_{ij}\in C$, where $C$ is a discrete non-negative random variable that represents latent censoring times. Assumed to be independent from $T$, this $C$ has the survivor function $G(u) = \mathbb{P}(C > u)$, along with a corresponding estimate $\hat{G}(u)$ using the Kaplan-Meier estimator.
Furthermore, let $\delta_{ij}=\mathbb{I}\left(T_{ij}\leq C_{ij}\right) \in \{0,1\}$ be the overall event indicator such that $\delta_{ij}=1$ if the event is observed at some point, and 0 otherwise. Lastly, we observe the overall resolution time $\tilde{T}_{ij}=\min{(T_{ij},C_{ij})}$.

\citet{Graf1999} explained that there are three categories of contributions to the tBS, each with its own weight in accounting for the loss of information due to censoring. Firstly, those subject-spells that have experienced the event by a given $t^*$, i.e., $\tilde{T}_{ij}\leq t^*$ and $\delta_{ij}=1$, are weighed by $1/\hat{G}\left(\tilde{T}_{ij}\right)$. In this case, the tBS-contribution is $(0-\hat{p}_{ij}(t^*))^2$ since $Y_{ij}(t^*)=0$. Secondly, those who have survived beyond $t^*$, i.e., $\tilde{T}_{ij}>t^*$ regardless of $\delta_{ij}$, have a tBS-contribution of $(1-\hat{p}_{ij}(t^*))^2$, and are weighed by $1/\hat{G}\left(t^*\right)$ since they are still at-risk. For the censored cases, i.e., $\tilde{T}_{ij}\leq t^*$ with $\delta_{ij}=0$, one cannot calculate the tBS-contribution and these cases are therefore weighed by 0. \citet{Gerds2006} called this weighting scheme the \textit{inverse probability of censoring} (IPC) weighting-method in producing an unbiased estimator of tBS. Accordingly, and for subjects $i=1,\dots,N_p$ with spells $j=1,\dots,n_i$, the IPC-weighted estimator of the tBS at $t^*$ is 
\begin{equation} \label{eq:tBS_IPCW}
    \widehat{\mathrm{tBS}}(t^*) = \frac{1}{\sum_{i=1}^{N_p} n_i} \sum_{i=1}^{N_p} \sum_{j=1}^{n_i} \left[
    \frac{\left(0-\hat{p}_{ij}\left(t^*\right)\right)^2 \cdot \mathbb{I}\left(\tilde{T}_{ij} \leq t^*; \delta_{ij}=1 \right)}{\hat{G}\left( \tilde{T}_{ij} \right)} + 
    \frac{\left(1-\hat{p}_{ij}\left(t^*\right)\right)^2 \cdot \mathbb{I}\left( \tilde{T}_{ij}>t^* \right)}{\hat{G}\left( t^* \right)}
    \right] \, .
\end{equation}

In \autoref{eq:tBS_IPCW}, only those subject-spells that are uncensored at $t^*$ can contribute their survival probabilities $\hat{p}_{ij}(t^*)$ to $\widehat{\mathrm{tBS}}(t^*)$. 
More broadly, \citet{suresh2022survival} noted that a key advantage of the tBS is its \textit{model-agnostic} nature: it does not assume correctness of the underlying survival model and only relies on predicted survival probabilities. The IPC-weighting approach further avoids the need to model the relationship between covariates and event times explicitly, as explained by \citet{Gerds2006}.
To summarise predictive performance across a wider time horizon $[0, t^*]$, the \textit{integrated Brier score} (IBS) is computed as
\[
\mathrm{IBS}(0, t^*) = \int_0^{t^*} \mathrm{tBS}(s) \, dW(s),
\]
where $W(s)$ is some weighting function. In the interest of simplicity, we have chosen a uniform weight, i.e., $W(s)=1/t^*$. Overall, the IBS provides a single scalar summary of predictive error across time, analogous to the integrated area under the tROC-graph. Both the tBS and the IBS are implemented in the \texttt{riskRegression}-package in the R-programming language, as maintained by \citet{riskRegression}. However, we also developed a bespoke implementation of the tBS and IBS, which is compiled in script 0e and applied in script 6d within the codebase by \citet{botha2025termStructureSourcecode}. Our \texttt{tBrierScore()}-function uniquely caters for left-truncated spells in the counting process style, as well as for recurrent default events within a multi-spell setup.

%% file: 7-InputSpace.tex
\section{Appendix: The input spaces of two competing discrete-time hazard models}
\label{app:InputSpace}

In \autoref{tab:featuresDescription}, we list and describe the various input variables that were used in our two discrete-time hazard (Dth) models. These input variables include account-level variables that span both time-fixed application and time-varying behavioural variables, as well as portfolio-level and macroeconomic variables.
The various coefficient estimates are deemed as confidential, though most of the remaining details are relegated in the interest of brevity to the open-source R-codebase maintained by \citet{botha2025termStructureSourcecode}. All of these variables are deemed as statistically significant at the significance level of 5\%. 
Regarding variable selection, we follow a \textit{thematic selection process} that is based on the interplay amongst domain knowledge, statistical analyses, and expert judgment. Themed questions guide the investigation; e.g., \textit{"which lag-order of a particular macroeconomic variable is best?"}, which ultimately produce insights on the `best' variable.
For each themed question, repeated logistic regressions are conducted and the resulting models are statistically analysed using various diagnostics, e.g., statistical significance, Akaike's Information Criterion (AIC), McFadden's pseudo $R^2$, and classical ROC-analyses.
Correlation studies and stepwise forward selection are also used at various points of this process, mainly towards combining insights from previous analyses. Lastly, a single-factor model is built for each of the curated variables towards establishing variable importance, whereafter the models (and underlying variables) are ranked in terms of the AIC and concordance statistics.
Details hereof are given in the accompanying codebase.

\begin{longtable}{p{3.7cm} p{9.3cm} p{1.3cm}}
\caption{The selected input variables, mapped across the two DtH-models. Subscripts $[\mathrm{a}]$ denote loan account-level variables, $[\mathrm{p}]$ are portfolio-level inputs, and $[\mathrm{m}]$ represent macroeconomic covariates.}
\label{tab:featuresDescription} \\
\toprule
\textbf{Variable} & \textbf{Description} & \textbf{Models} \\ 
\midrule
\endfirsthead
\caption[]{(continued)} \\
\toprule
\textbf{Variable} & \textbf{Description} & \textbf{Models} \\ 
\midrule
\endhead
\midrule \multicolumn{3}{r}{\textit{Continued on next page}} \\
\endfoot
\bottomrule
\endlastfoot
\footnotesize{\texttt{AgeToTerm\_Avg}$_{[\mathrm{p}]}$}  & \footnotesize{Mean value across the portfolio of the ratio between a loan's age and its term.} & \footnotesize{Advanced} \\
\footnotesize{\texttt{ArrearsDir\_3\_Changed}$_{[\mathrm{a}]}$}  & \footnotesize{Boolean variable indicating whether a change occurred in the trending direction of the arrears balance over 3 months. This direction is obtained qualitatively by comparing the current arrears-level to that of 3 months ago, binned as: 1) increasing; 2) milling; 3) decreasing (reference); and 4) missing.} & \footnotesize{Advanced} \\
\footnotesize{\texttt{Arrears}$_{[\mathrm{a}]}$} & \footnotesize{Amount in arrears.} & \footnotesize{Basic} \\
\footnotesize{\texttt{g0\_Delinq\_1}$_{[\mathrm{a}]}$}  & \footnotesize{1-month lagged version of the $g_0$ delinquency measure (\texttt{g0\_Delinq}) from \citet{botha2021paper1}, which evaluates the number of payments in arrears at a point in time.} & \footnotesize{Advanced}\\
\footnotesize{\texttt{g0\_Delinq\_SD\_4}$_{[\mathrm{a}]}$} & \footnotesize{The sample standard deviation of \texttt{g0\_Delinq} over a rolling 4-month window.} &  \footnotesize{Advanced} \\
\footnotesize{\texttt{InterestRate\_Nominal}$_{[\mathrm{a}]}$} & \footnotesize{Nominal interest rate per annum of a loan.} & \footnotesize{Advanced, Basic} \\
\footnotesize{\texttt{InstalToBalance\_Sum}$_{[\mathrm{p}]}$} & \footnotesize{The sum of instalments across the portfolio divided by the sum of outstanding balances.} & \footnotesize{Advanced} \\
\footnotesize{\texttt{M\_DebtToIncome}$_{[\mathrm{m}]}$}  &\footnotesize{Debt-to-Income: Average household debt expressed as a percentage of household income per quarter, interpolated monthly.} & \footnotesize{Advanced} \\
\footnotesize{\texttt{M\_Inflation\_Growth\_6}$_{[\mathrm{m}]}$} & \footnotesize{Year-on-year growth rate in inflation index (CPI) per month, lagged by six months.} & \footnotesize{Advanced, Basic} \\ 
\footnotesize{\texttt{NewLoans\_Pc}$_{[\mathrm{p}]}$} & \footnotesize{Fraction of the portfolio that constitutes new loans.} & \footnotesize{Advanced} \\
\footnotesize{\texttt{PayMethod}$_{[\mathrm{a}]}$} & \footnotesize{A categorical variable designating different payment methods: 1) debit order (reference); 2) salary; 3) payroll or cash; and 4) missing.} & \footnotesize{Advanced} \\
\footnotesize{\texttt{RollEver\_24}$_{[\mathrm{a}]}$} & \footnotesize{Number of times that loan delinquency increased during the last 24 months, excluding the current time point.} & \footnotesize{Advanced} \\
\footnotesize{\texttt{SpellNum\_Bn}$_{[\mathrm{a}]}$} & \footnotesize{The current performance spell number, or total number of visits across all spells over loan life, binned as ("1", "2", "3", "4+") spells.} & \footnotesize{Advanced, Basic} \\ 
\footnotesize{\texttt{Time\_Bn}$_{[\mathrm{a}]}$} & \footnotesize{A binned version of the time spent in a performing spell towards embedding the baseline hazard.} & \footnotesize{Advanced, Basic} \\ 
\footnotesize{\texttt{Time\_Bn*SpellNum\_Bn}$_{[\mathrm{a}]}$} & \footnotesize{An interaction effect between \texttt{Time\_Bn} and \texttt{SpellNum\_Bn}.} & \footnotesize{Advanced} \\
\footnotesize{\texttt{TimeSpell*SpellNum\_Bn}$_{[\mathrm{a}]}$} & \footnotesize{An interaction effect between the logarithm of the time spent in a performing spell, and \texttt{SpellNum\_Bn}.} & \footnotesize{Basic} \\
\end{longtable}